\documentclass[oldversion]{aa}  
\usepackage{graphicx,amsmath}
\usepackage{amssymb}
\usepackage{color}
\usepackage{natbib}             
\usepackage{url}
\usepackage{multirow}
\usepackage{threeparttable}
\bibpunct{(}{)}{;}{a}{}{,} 
\usepackage{txfonts}
\usepackage{subfig}

\def\approxinf{%
  \def\p{%
    \setbox0=\vbox{\hbox{$<$}}%
    \ht0=0.6ex \box0 }%
  \def\s{%
    \vbox{\hbox{$\sim$}}%
  }%
  \mathrel{\raisebox{0.7ex}{%
      \mbox{$\underset{\s}{\p}$}%
    }}%
}

\def\approxsup{%
  \def\p{%
    \setbox0=\vbox{\hbox{$>$}}%
    \ht0=0.6ex \box0 }%
  \def\s{%
    \vbox{\hbox{$\sim$}}%
  }%
  \mathrel{\raisebox{0.7ex}{%
      \mbox{$\underset{\s}{\p}$}%
    }}%
}

\begin{document}

   \title{High-energy environment of super-Earth 55\,Cnc e I: Far-UV chromospheric variability as a possible tracer of planet-induced coronal rain\thanks{Observational templates of the geocoronal emission are available in electronic form at the CDS via anonymous ftp to cdsarc.u-strasbg.fr (130.79.128.5) or via http://cdsweb.u-strasbg.fr/cgi-bin/qcat?J/A+A/}}
                                   
   \author{
   V.~Bourrier\inst{1},
   D.~Ehrenreich\inst{1},
   A.~Lecavelier des Etangs\inst{2},
   T.~Louden\inst{3},  
   P.J.~Wheatley\inst{3},
   A.~Wyttenbach\inst{1},
   A.~Vidal-Madjar\inst{2},
   B.~Lavie\inst{1},
   F.~Pepe\inst{1},   
   S.~Udry\inst{1}                  
        }
   
\authorrunning{V.~Bourrier et al.}
\titlerunning{High-energy environment of super-Earth 55\,Cnc e. I}

\offprints{V.B. (\email{vincent.bourrier@unige.ch})}

\institute{
Observatoire de l'Universit\'e de Gen\`eve, 51 chemin des Maillettes, 1290 Sauverny, Switzerland
\and 
Institut d'astrophysique de Paris, UMR7095 CNRS, Universit\'e Pierre \& Marie Curie, 98bis boulevard Arago, 75014 Paris, France 
\and   
Department of Physics,University of Warwick, Coventry CV4 7AL, UK
}

   \date{} 
 
  \abstract
{The high-energy X-ray to ultraviolet (XUV) irradiation of close-in planets by their host star influences their evolution and might be responsible for the existence of a population of ultra-short period planets eroded to their bare core. In orbit around a bright, nearby G-type star, the super-Earth 55\,Cnc e offers the possibility to address these issues through transit observations at UV wavelengths. We used the Hubble Space Telescope to observe the transit in the far-ultraviolet (FUV) over three epochs in April 2016, January 2017, and February 2017. Together, these observations cover nearly half of the orbital trajectory in between the two quadratures, and reveal significant short- and long-term variability in 55\,Cnc chromospheric emission lines. In the last two epochs, we detected a larger flux in the \ion{C}{iii}, \ion{Si}{iii,} and \ion{Si}{iv} lines after the planet passed the approaching quadrature, followed by a flux decrease in the \ion{Si}{iv} doublet. In the second epoch these variations are contemporaneous with flux decreases in the \ion{Si}{ii} and \ion{C}{ii} doublet. All epochs show flux decreases in the \ion{N}{v} doublet as well, albeit at different orbital phases. These flux decreases are consistent with absorption from optically thin clouds of gas, are mostly localized at low and redshifted radial velocities in the star rest frame, and occur preferentially before and during the planet transit. These three points make it unlikely that the variations are purely stellar in origin, yet we show that the occulting material is also unlikely to originate from the planet. We thus tentatively propose that the motion of 55 Cnc e at the fringes of the stellar corona leads to the formation of a cool coronal rain. The inhomogeneity and temporal evolution of the stellar corona would be responsible for the differences between the three visits. Additional variations are detected in the \ion{C}{ii} doublet in the first epoch and in the \ion{O}{i} triplet in all epochs with a different behavior that points toward intrinsic stellar variability. Further observations at FUV wavelengths are required to disentangle definitively between star-planet interactions in the 55\,Cnc system and the activity of the star.}

\keywords{planetary systems - Stars: individual: 55\,Cnc - methods: data analysis - techniques: spectroscopic - planets and satellites: individual: 55\,Cnc e - stars: chromospheres - ultraviolet: stars}

   \maketitle

\section{Introduction}
\label{intro} 

The population of close-in planets is shaped by interactions with their host star. In particular, the deposition of stellar X-ray and extreme ultraviolet radiation (XUV) into an exoplanet upper atmosphere can lead to its hydrodynamic expansion and the escape of large amounts of gas from the gravitational well of the planet (e.g., \citealt{VM2003}, \citealt{Lecav2004}, \citealt{Johnstone2015}). This evaporation can sustain extended exospheres that have been detected around Jupiter-mass planets (\citealt{VM2003,VM2004}; \citealt{Lecav2010,Lecav2012}, \citealt{Bourrier2013}; \citealt{Ehrenreich2012}; \citealt{Fossati2010}, \citealt{Haswell2012}) and a Neptune-mass planet (\citealt{Kulow2014}; \citealt{Ehrenreich2015}, \citealt{Lavie2017}). While Jupiter-mass planets are too massive to be significantly affected by evaporation, losing a few percent of their mass over their lifetime (e.g., \citealt{Lecav2007}, \citealt{Hubbard2007}, \citealt{Ehrenreich_desert2011}), lower mass planets could be stripped of most of their atmosphere (e.g., \citealt{Lecav2007}, \citealt{owen2012}) and evolve into chthonian planets (\citealt{Lecav2004}). There is a desert in the population of close-in planets that was first identified as a lack of sub-Jupiter planets (e.g., \citealt{Lecav2007}, \citealt{Davis2009},  \citealt{szabo_kiss2011}, \citealt{beauge2013}) and later shown to extend to strongly irradiated super-Earths (\citealt{Lundkvist2016}). Theoretical studies show that this desert is explained well by planets with gaseous envelopes large enough to capture much of the stellar energy and evaporate, but too light to retain their escaping atmospheres (e.g., \citealt{Lopez2012,Lopez2013}, \citealt{Owen2013}, \citealt{Kurokawa2014}, \citealt{Jin2014}). Recently, \citet{Fulton2017} have identified a deficit of small close-in planets with radii of about 1.7\,R$_\oplus$ separating two peaks in the radius distribution at about 1.3 and 2.5\,R$_\oplus$. There seems to be a dichotomy between large super-Earths massive enough to retain H/He envelopes with mass fractions of a few percent, and small rocky super-Earths with atmospheres that contribute negligibly to their size (\citealt{Weiss2014}; \citealt{Rogers2015}). The absence of planets in between  (the so-called evaporation valley) would arise because the planetary radius barely changes as the envelope mass decreases, resulting in more tenuous atmospheres receiving the same amount of irradiation, and thus more prone to escape (\citealt{Owen2017}).\\

Understanding the evolution of super-Earths subjected to atmospheric escape requires that we probe the upper atmosphere of planets on both sides of the valley. On the lower radius side of the valley, transit observations at Ly-$\alpha$ wavelengths have revealed hints of hydrogen exospheres from the outer warm sub-Earths around the ancient star Kepler-444 (\citealt{Bourrier2017_K444}), while a similar search is ongoing for the seven Earth-size temperate planets around the ultra-cool dwarf TRAPPIST-1 (\citealt{Bourrier2017_recoT1}, \citealt{Bourrier2017_waterT1}). On the upper-radius side of the valley, Ly-$\alpha$ observations have revealed no signatures of neutral hydrogen escape from the mildly irradiated super-Earth HD\,97658 b (\citealt{Bourrier2017_HD976}) and the highly irradiated 55 Cnc e (\citealt{Ehrenreich2012}). 55\,Cnc e orbits its star in a mere 17.7\,hours and is a member of an extreme population of planets with ultra-short periods (USPs; $P\approxinf$1\,day). With a radius of 1.9\,R$_\oplus$, 55\,Cnc e is one of the largest USPs and stands at the edge of the upper-radius side of the evaporation valley. These features, and its presence within a compact multi-planets system, raises question about the formation and evolution of 55\,Cnc e under the combined influence of its star and companion planets, and about the existence of a putative atmosphere. \\

\begin{center}
\begin{figure}[tbh!]
\centering
\includegraphics[trim=0.4cm 0cm 0cm 0.1cm,clip=true,width=0.9\columnwidth]{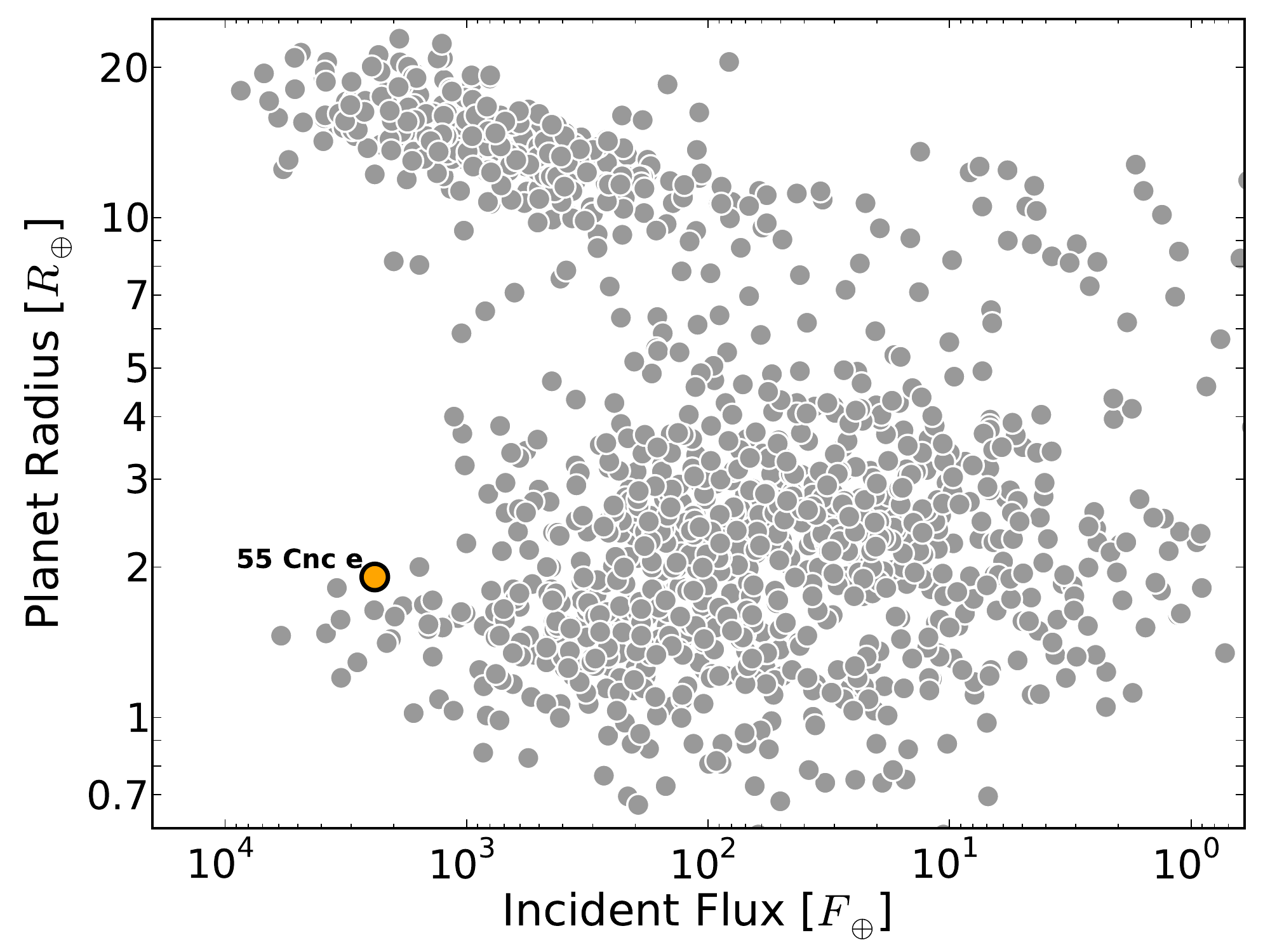}
\caption[]{Distribution of planetary radius as a function of bolometric incident stellar flux, expressed in units of flux received at Earth. Physical properties of exoplanets were extracted from the Extrasolar Planets Encyclopaedia (exoplanet.eu) in November 2016. Evaporation shaped the population of hot close-in planets, likely resulting in the formation of a desert of Neptune and a valley of super-Earths stripped of their atmospheres (see text). 55\,Cnc e (orange disk) stands on the upper side of the valley. }
\label{fig:Radius_flux}
\end{figure}
\end{center}

The planet 55 Cnc e (designated Janssen according to the IAU; \citealt{Montmerle2016}) was first detected and characterized with radial velocity measurements (\citealt{Fischer2008}, \citealt{Dawson2010}), then later observed in transit with MOST optical (\citealt{Winn2011}) and Spitzer infrared observations (\citealt{demory2011}). Subsequent photometric and velocimetric follow-up refined the planet properties (\citealt{Demory2012}; \citealt{Gillon2012}; \citealt{Endl2012}, \citealt{Dragomir2013}, \citealt{deMooij2014}, \citealt{Demory2016a}). Because 55\,Cnc e transits the bright (V=6) and nearby (d=13\,pc) star Copernicus, it has been one of the most promising super-Earths for atmospheric characterization. The strong irradiation from its G8 host star (\citealt{Ehrenreich2012}, \citealt{Demory2016a}) makes it unlikely to harbor a substantial H/He envelope, but could allow for a water or carbon-rich envelope that is resilient to atmospheric escape (\citealt{Lopez2012}, \citealt{tian2009}). Initial estimates of 55\,Cnc e bulk density were further consistent with a silicate-rich interior and water envelope (\citealt{demory2011}; \citealt{Winn2011}; \citealt{Gillon2012}) or a carbon-rich interior and no envelope (\citealt{Madhusudhan2012}). The absence of a hydrogen exosphere around 55\,Cnc e (\citealt{Ehrenreich2012}) and the non-detection of water through ground-based optical spectroscopy (\citealt{Esteves2017}) suggest that the planet does not harbor an extended water-rich envelope. \\
Infrared observations of 55\,Cnc e revealed a high temporal variability over timescales of a few months, in the form of significant changes in the occultation depth and dayside thermal emission (\citealt{Demory2012}, \citealt{Demory2016a}), and marginal variations in its transit depth (\citealt{Demory2016a}). Observations of the phase curve and occultations in the infrared (\citealt{Demory2016b}) revealed that the planet presents a hot spot about 40$^{°}$ degrees east of the substellar point, and inefficiently redistributes heat from its dayside (2700$\pm$270\,K) to its nightside (1380$\pm$400\,K). Despite the claim of an atmospheric signature from HCN by \citet{Tsiaras2016}, these infrared observations thus confirm that 55\,Cnc e does not harbor a light H/He envelope, and suggest that the planet either has an optically thick atmosphere with heat recirculation confined to the planetary dayside or no atmosphere but low-viscosity magma flows at its surface. Further analysis of 55\,Cnc e infrared phase curve favors a scenario with a substantial, heavy-weight atmosphere (\citealt{Angelo2017}). Analyzing the composition and structure of metals in the exosphere around 55\,Cnc e would allow us to disentangle between the magmatic and atmospheric scenarios. High-resolution optical transit observations have revealed intriguing variations in the sodium and singly-ionized calcium lines, possibly arising from an exosphere exhibiting high temporal variability (\citealt{RiddenHarper2016}). \\
Observations at shorter wavelengths in ultraviolet (UV) lines could reveal the presence of further species in the exosphere of 55\,Cnc e, such as ionized carbon or silicon. Measurements at UV wavelengths can also trace interactions between the planet and the stellar chromosphere. Signatures of star-planet interactions (SPIs) usually relate to an enhancement in the activity of the star (photospheric, chromospheric, or coronal) at specific planetary orbital  phases (\citealt{Shkolnik2003,Shkolnik2005,Shkolnik2013}; \citealt{Saar2008}, \citealt{Kashyap2008}, \citealt{Walker2008}; \citealt{Pillitteri2010,Pillitteri2015}; \citealt{Lanza2011}), which are attributed to heating from tidal or magnetic interactions with the planet or to planetary material accreting onto the star (see \citealt{Cuntz2000}, \citealt{Strugarek2017}). Conversely, abnormally low flux in the core of stellar chromospheric lines has been attributed to material escaped from close-in planets, condensing in a circumstellar torus at a larger orbital distance or closer-in within the coronal field (\citealt{Haswell2012}; \citealt{Lanza2014}, \citealt{Fossati2015}). Finally, measuring the UV emission of 55\,Cnc is essential to determining the radiative environment of all planets in the system, which is a key input to study the physicochemical processes in their atmosphere. We thus obtained FUV and near-ultraviolet (NUV) transit observations of 55\,Cnc e with the Hubble Space Telescope (HST) at multiple epochs to study the planet and its environment. Owing to the large number of datasets and their different spectral domains and reduction techniques, we decided to make a series of publications focusing on complementary science cases. In this paper (paper I) we investigate the temporal variability of the stellar emission lines observed at three epochs in the FUV. The FUV transit observations and their reduction are presented in Sect.~\ref{sec:obs}. In Sect.~\ref{sec:ttt} we search for spectro-temporal variations in the brightest lines available in the observed spectral range. Signatures of variability are interpreted in Sect.~\ref{sec:SPIs}. We conclude this first study in Sect.~\ref{sec:conclu}.\\

\section{Observations and data reduction}
\label{sec:obs} 

\subsection{Description}
\label{sec:obs_desc}

We observed 55\,Cnc in the FUV domain with the Cosmic Origin Spectrograph (COS) instrument on board the HST. We used the medium resolution G130M grating centered at 1291\,\AA\,, which covers the range 1135--1432\,\AA\, with two independent spectra separated by a 14.3\,\AA\, gap at $\sim$1280\,\AA. As a cool G8 star, 55\,Cnc has no continuum emission at these wavelengths, and only coronal and chromospheric emission lines provide a background source to perform transit observations. We focused our analysis on the brightest lines in the grating range, given in Table~\ref{tab:studied_lines}. Scientific exposures were obtained by scanning the four positions available on the COS detector, using the FP-POS=ALL setting. Each FP-POS$_{i}$ sub-exposure (hereafter referred to as FP$_{i}$ exposure, where $i$ = 1 to 4) is slightly offset in the dispersion direction, causing spectral features to fall on a different part of the detector and reducing the impact of fixed-pattern noise in the COS detectors. The CALCOS pipeline (version 3.1.7) creates calibrated spectra for each FP$_{i}$, then aligns and combines these spectra into a merged spectral product that contains only good-quality data at each wavelength and no gap. These orbit-long spectra were used for all analyses hereafter, except for the study of the oxygen lines (Sect. ~\ref{sec:spec_line}). We note that scanning different positions on the detector likely limited the stability of the spectral calibration (Sect.~\ref{sec:flux_w_cal}), therefore we do not recommend this strategy for future COS FUV observations.\\
 
\begin{table}[h!]
\caption[]{Stellar lines used in our analysis}
\centering
\begin{threeparttable}
\begin{tabular}{lcc}
\hline
\hline
Transition & Wavelength (\,\AA) & $\log\,T_\mathrm{max}$ (K) \\
\hline
\ion{C}{iii} & 1174.93 &  4.8   \\
 & 1175.26 &    \\
 & 1175.59 &   \\
 & 1175.71 &    \\
 & 1175.99 &   \\ 
 & 1176.37 &   \\ 
\ion{Si}{iii} & 1206.50 & 4.7    \\ 
\ion{O}{v} & 1218.344 & 5.3   \\ 
\ion{N}{v} & 1238.821     & 5.2  \\
          & 1242.804 &    \\
\ion{Si}{ii} & 1264.738 & 4.2  \\ 
                         & 1265.002 &  \\ 
\ion{O}{i} & 1302.168 &  3.9  \\ 
           & 1304.858 &   \\ 
           & 1306.029 &   \\
\ion{C}{ii} & 1334.532 &  4.5  \\
            & 1335.708 &    \\
\ion{Si}{iv} & 1393.755   & 4.9  \\
             & 1402.770 &  \\
    \hline
  \end{tabular}
  \begin{tablenotes}[para,flushleft]
  Notes: Second column indicates rest wavelengths from the NIST Atomic Spectra Database (\citealt{Kramida2016}). Formation temperatures are from the Chianti v.7.0 database (\citealt{Dere1997}, \citealt{Landi2012}). 
  \end{tablenotes}
  \end{threeparttable}
\label{tab:studied_lines}
\end{table}

\begin{center}
\begin{figure}[tbh!]
\centering
\includegraphics[trim=0cm 0cm 0cm 0cm,clip=true,width=0.8\columnwidth]{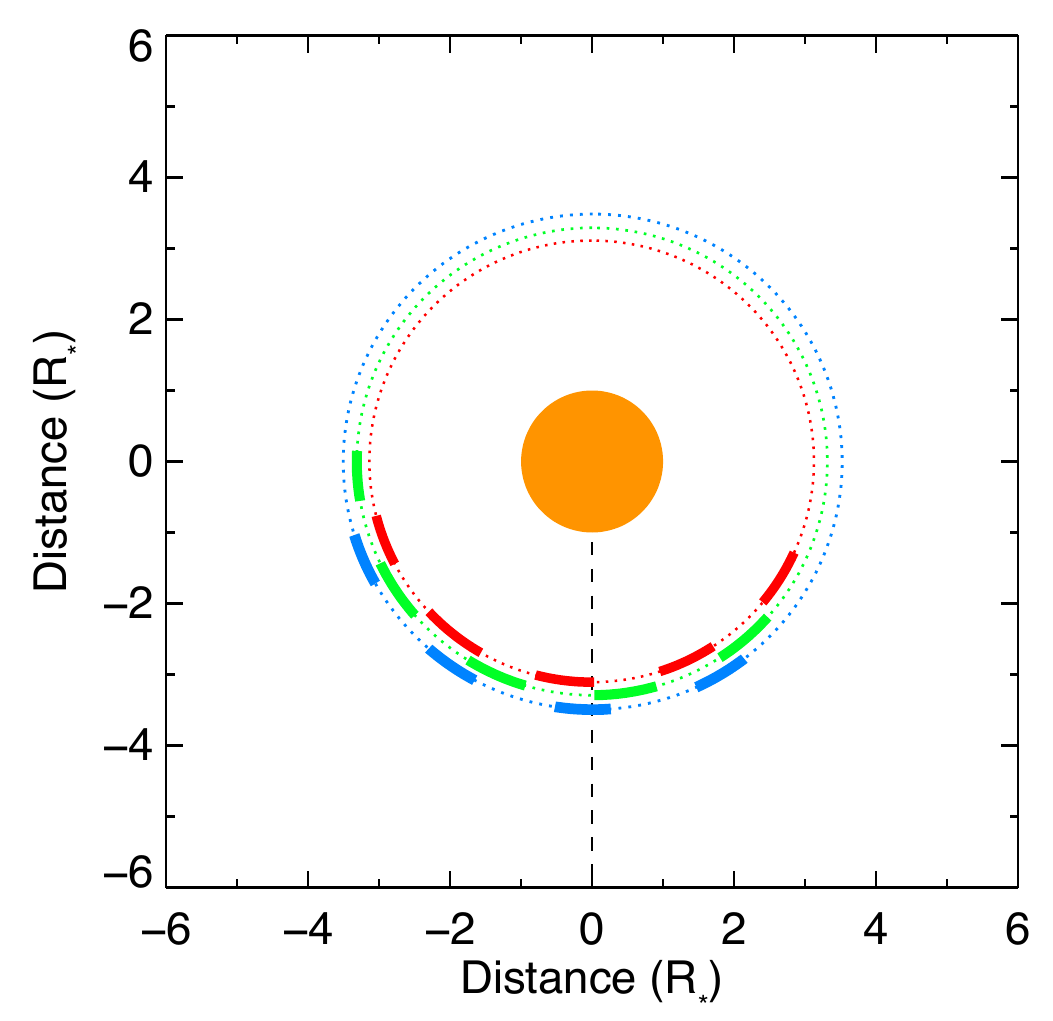}
\caption[]{Orbital positions of 55\,Cnc e at the time of the HST observations in visits A$_\mathrm{fuv}$ (blue), B$_\mathrm{fuv}$ (green), and C$_\mathrm{fuv}$ (red). Each rectangle corresponds to the space covered by the planet during one of the HST orbits. The planetary system is viewed from above, and 55\,Cnc e is moving counterclockwise. The dashed black line indicates the line of sight toward Earth. Star and orbital trajectory in Visit B$_\mathrm{fuv}$ have the correct relative scale, while the semimajor axis was adjusted in visits A$_\mathrm{fuv}$ and C$_\mathrm{fuv}$ for the sake of clarity.}
\label{fig:Couv_orbitale}
\end{figure}
\end{center}

After a first visit revealed significant variations in several stellar lines (GO Program 14094, PI: V. Bourrier), two additional visits were granted to investigate their origin (Mid-Cycle Program 14877; PI: V. Bourrier). The log of the three visits is given in Table~\ref{tab:log}. Four HST orbits were used in Visit A$_\mathrm{fuv}$ (April 4, 2016), while five HST orbits were used in Visit B$_\mathrm{fuv}$ (January 5, 2017) and in Visit C$_\mathrm{fuv}$ (February 14, 2017). Scientific exposures range between $\sim$29 and 37\,min for each orbit (depending on the visit and the orbit) and were obtained at different orbital phases relative to the transit of 55\,Cnc e (Fig.~\ref{fig:Couv_orbitale}). Nonetheless all visits have exposures before, during, and after the optical transit (see for example Fig.~\ref{fig:LC_CIII_SiIII_SiIV}). We set the orbital and bulk properties of 55\,Cnc e to the values derived by \citet{Demory2016b} and used the ephemeris derived from a combined fit to all available Spitzer data (T$_{0}$=2455733.0059$\stackrel{+0.0010}{_{-0.0015}}$\,BJD$_\mathrm{TDB}$, P=0.7365463$\stackrel{+0.0000011}{_{+0.0000018}}$\,days; private communication from B.O.~Demory). The radial velocity of the star was set to 27.58\,km\,s$^{-1}$ (\citealt{Nidever2002}). Although the 55\,Cnc system hosts at least five planets, only 55\,Cnc e transits the host star in the optical. The warm giant 55\,Cnc b could have an extended atmosphere of neutral hydrogen that undergoes a partial transit in the stellar Ly-$\alpha$ line (\citealt{Ehrenreich2012}), and which could potentially yield similar transit signatures in other FUV lines. We ensured that 55\,Cnc b was far enough from inferior conjunction during each visit (about -130$^{\circ}$ in Visit A$_\mathrm{fuv}$, -180$^{\circ}$ in Visit B$_\mathrm{fuv}$, and 80$^{\circ}$ in Visit C$_\mathrm{fuv}$) so that it did not contaminate our observations. \\

\begin{table}[tbh]
\centering
\caption{Log of 55\,Cnc e transit observations.}
\begin{tabular}{lccc}
\hline
\hline
\noalign{\smallskip}
Visit & Date & \multicolumn{2}{c}{Time (UT)}   \\
      &      & Start & End                                \\
\noalign{\smallskip}
\hline
A$_\mathrm{fuv}$ & 4 April 2016     & 09:19:36   & 14:37:10 \\
B$_\mathrm{fuv}$ & 5 January 2017     & 13:12:05   &  20:07:00\\
C$_\mathrm{fuv}$ & 14 February 2017      & 08:35:53  & 15:33:45\\
\noalign{\smallskip}
\hline
\hline
\end{tabular}
\label{tab:log}
\end{table}

Because of the ultra-short period of 55\,Cnc e (17.7\,h), our three visits cover nearly 180$^{\circ}$ of the planet revolution from the approaching to near the receding quadrature (Fig.~\ref{fig:Couv_orbitale}). Over the course of our observations, the radial velocity of the planet in the star rest frame increases from about -230 to 210\,km\,s$^{-1}$. Even over a single HST exposure, the planet radial velocity varies by up to 50\,km\,s$^{-1}$ (for mid-transit exposures), which means that absorption signals from a putative exosphere are expected to be blurred over a large spectral range. \\

\subsection{Flux uncertainties and spectral calibration}
\label{sec:flux_w_cal}

Similar to \citet{Wilson2017}, we found that the uncertainties produced by the CALCOS pipeline on the flux values were largely overestimated compared to the dispersion of the measurements, which was found to be consistent with a noise budget dominated by photon noise. We thus calculated all uncertainties using Equation (1) in \citet{Wilson2017}. \\

\begin{center}
\begin{figure}[h!]
\centering
\includegraphics[trim=1.7cm 6cm 8.5cm 6cm,clip=true,width=\columnwidth]{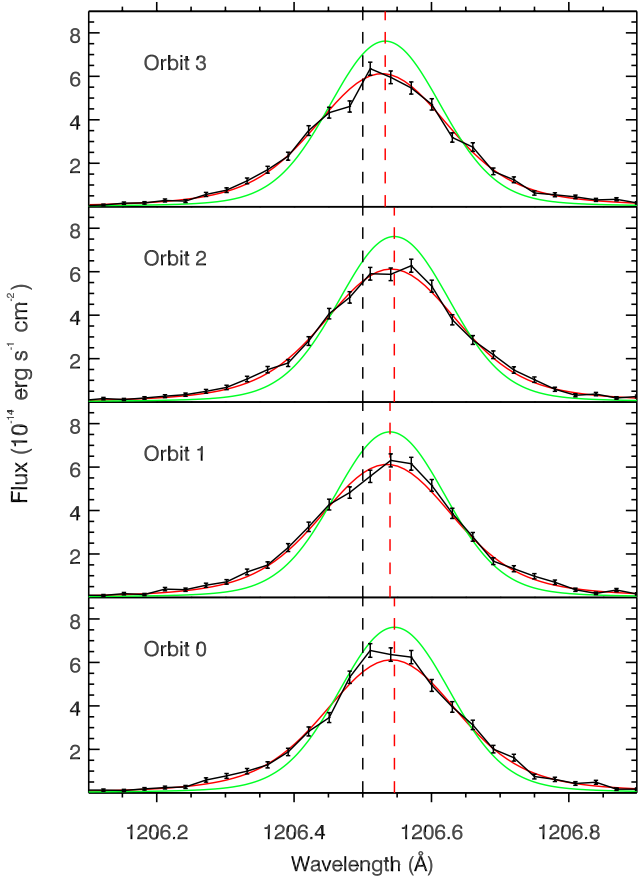}
\caption[]{\ion{Si}{iii} line in Visit A$_\mathrm{fuv}$, plotted as a function of wavelength in the expected star rest frame. The green line shows the theoretical profile for the intrinsic stellar line, which is convolved with COS LSF (yielding the red spectra) to be compared to the observations (black spectra). The dashed black line shows the expected rest wavelength of the stellar line, and the dashed red line its measured central wavelength. We note that COS LSFs are asymmetrical and off-centered, hence the observed stellar lines are slightly distorted compared to the intrinsic lines.   }
\label{fig:SiIII_V2_calib}
\end{figure}
\end{center}

The wavelength tables of the spectra were corrected for the radial velocity of 55\,Cnc, which we take to be representative of its photosphere. We then noticed significant spectral shifts between the expected rest wavelength of the stellar lines and their actual position. These shifts vary between the orbits of a given visit in the range of up to $\sim$5\,km\,s$^{-1}$ and the average shift within a visit ranges between about -5\,km\,s$^{-1}$ and +25\,km\,s$^{-1}$, depending on the stellar line and its position on the detector (see for example Fig.~\ref{fig:SiIII_V2_calib}). Indeed, the shifts are a combination of biases in COS calibration (dependent on wavelength and varying with the visit) and the redshift of chromospheric emission lines (dependent on the stellar properties and formation temperature of the line; e.g., \citealt{Linsky2012}). A preliminary analysis of the raw spectra did not show extremely strong variations in the overall shape of the lines (see Fig.~\ref{fig:Lines_spec_grid_raw} in the Appendices). Therefore we assumed that variations could be quantified with respect to an average line profile, representative of the intrinsic stellar emission in each visit, which would remain at the same rest wavelength during a visit. Under this assumption, small variations in the line positions around this rest wavelength are either statistical or due to COS calibration bias, and we thus corrected the stellar lines for their full spectral shift measured in each exposure. This shift was measured by fitting each stellar line with analytical profiles, excluding spectral ranges showing possible variations. These ranges were identified by mirroring the lines in a given exposure and comparing them between various exposures (using the approach described in Sect.~\ref{sec:meth}) and by looking for deviations of the lines from preliminary best-fit analytical models. The intrinsic stellar lines were found to be well fitted with Gaussian or Voigt profiles, absorbed by the interstellar medium (ISM) where appropriate, and convolved with the non-Gaussian, slightly off-center COS LSF\footnote{LSF profiles were retrieved from the STScI website at \mbox{\url{http://www.stsci.edu/hst/cos/performance/spectral_resolution/}}, tabulated for the same settings as our observations (G130M grating at 1291\,\AA\, in Lifetime Position 3).}. All fits were performed on the spectra at COS sampling (about 0.01\,\AA), but for the sake of clarity all spectra displayed in this paper are binned by three pixels. We note that correcting the full spectral shifts (i.e., the combination of COS offsets and the chromospheric redshifts) biases the velocity range measured in the star rest frame for a line physically redshifted with respect to the photosphere. However the maximum redshift expected for a G8-type, slow-rotating star such as 55\,Cnc is about 5\,km\,s$^{-1}$ for the hottest chromospheric lines (P$_\mathrm{rot}\sim$53\,days; \citealt{Linsky2012}, \citealt{lopez2014}) and this offset does not change the conclusions of our study.

\subsection{Geocoronal contamination}
\label{sec:spec_line}

\subsubsection{Building airglow templates}

\begin{center}
\begin{figure}[tbh!]
\centering
\includegraphics[trim=2cm 6cm 2cm 12cm,clip=true,width=\columnwidth]{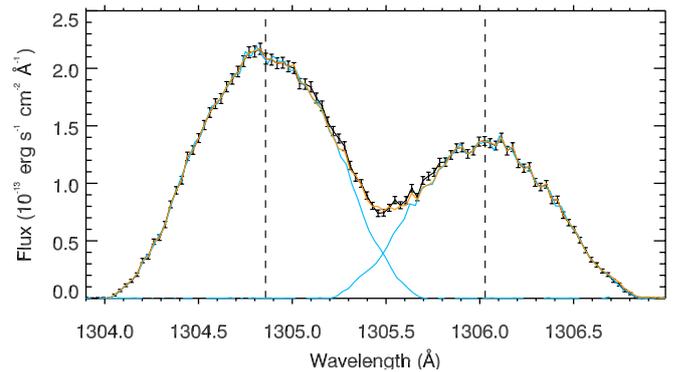}
\caption[]{Geocoronal emission in the excited \ion{O}{i}$_\mathrm{air}$ lines. The average of observed spectra (in black) was used to build individual templates for each line (in blue). The orange spectrum is the sum of the two templates, which fits well the blended parts of the lines.}
\label{fig:master_OI23_airglow}
\end{figure}
\end{center}

Geocoronal emission in our G130M spectra of 55\,Cnc contaminates the \ion{H}{i} stellar line and to a lesser extent the \ion{O}{i} triplet. In contrast to slit-based measurements with the STIS instrument, the circular aperture of COS prevents us from measuring the airglow independently from the stellar spectrum. However, the strength of the airglow varies significantly along an HST orbit. We found that the geocoronal emission in the oxygen triplet (\ion{O}{i}$_\mathrm{air}$ lines\footnote{Hereafter we use the subscript \textit{air} to designate geocoronal lines.}) is faint enough in most FP$_{i}$ such that the spectra show no detectable contamination. This concerns FP$_{2}$, $_{3}$, $_{4}$ in Visit A$_\mathrm{fuv}$, FP$_{1}$, $_{2}$, $_{3}$ in Visit B$_\mathrm{fuv}$, and all FP$_{i}$ in Visit C$_\mathrm{fuv}$. We defined airglow-free spectra of the stellar \ion{O}{i} lines by taking the mean of these sub-exposures in each orbit (Fig.~\ref{fig:Lines_spec_grid}). The same operation could not be carried for the stellar Ly-$\alpha$ line because it is always contaminated by the much brighter geocoronal \ion{H}{i}$_\mathrm{air}$ line. We thus assessed the possibility of correcting the stellar lines using observational airglow templates (e.g., \citealt{BJ_ballester2013}, \citealt{Wilson2017}), which we built from calibration observations of the airglow lines in the same grating as our observations\footnote{http://www.stsci.edu/hst/cos/calibration/airglow.html}. Our visits were obtained in lifetime position (LP) 3 (COS changed from LP2 to LP3 in February 2015), but we used all LP2/3 airglow observations available at the time of our study: four LP2 spectra from Program 13145, one LP3 spectrum from Program 13994, and four LP3 spectra from \citet{Wilson2017}. We shifted the airglow lines into their rest frame, and scaled them to the same flux level. After excluding sharp localized variations in some exposures, we found that all scaled lines have similar profiles and there are no significant differences between spectra obtained in LP2 or LP3. We thus coadded all observations to create templates of the \ion{H}{i}$_\mathrm{air}$ and \ion{O}{i}$_\mathrm{air}$ $\lambda$1302 lines. The same operation could not be carried out for the excited \ion{O}{i}$_\mathrm{air}$ lines because they are blended and their amplitude ratio changes from one epoch to another. However those lines are distant enough that the blue wing of the \ion{O}{i}$_\mathrm{air}$ $\lambda$1304 line and the red wing of the \ion{O}{i}$_\mathrm{air}$ $\lambda$1306 line are not blended, and we found no evidence for asymmetries in any of the \ion{O}{i}$_\mathrm{air}$ lines. Therefore we built individual airglow templates for the \ion{O}{i}$_\mathrm{air}$ $\lambda$1304 and \ion{O}{i}$_\mathrm{air}$ $\lambda$1306 lines by mirroring their unblended halves around each line center (Fig.~\ref{fig:master_OI23_airglow}). All airglow templates are made available in electronic form. \\

\subsubsection{Correcting for the airglow}
\label{sec:air_corr}

\begin{center}
\begin{figure}[tbh!]
\centering
\includegraphics[trim=0cm 0cm 0cm 0cm,clip=true,width=\columnwidth]{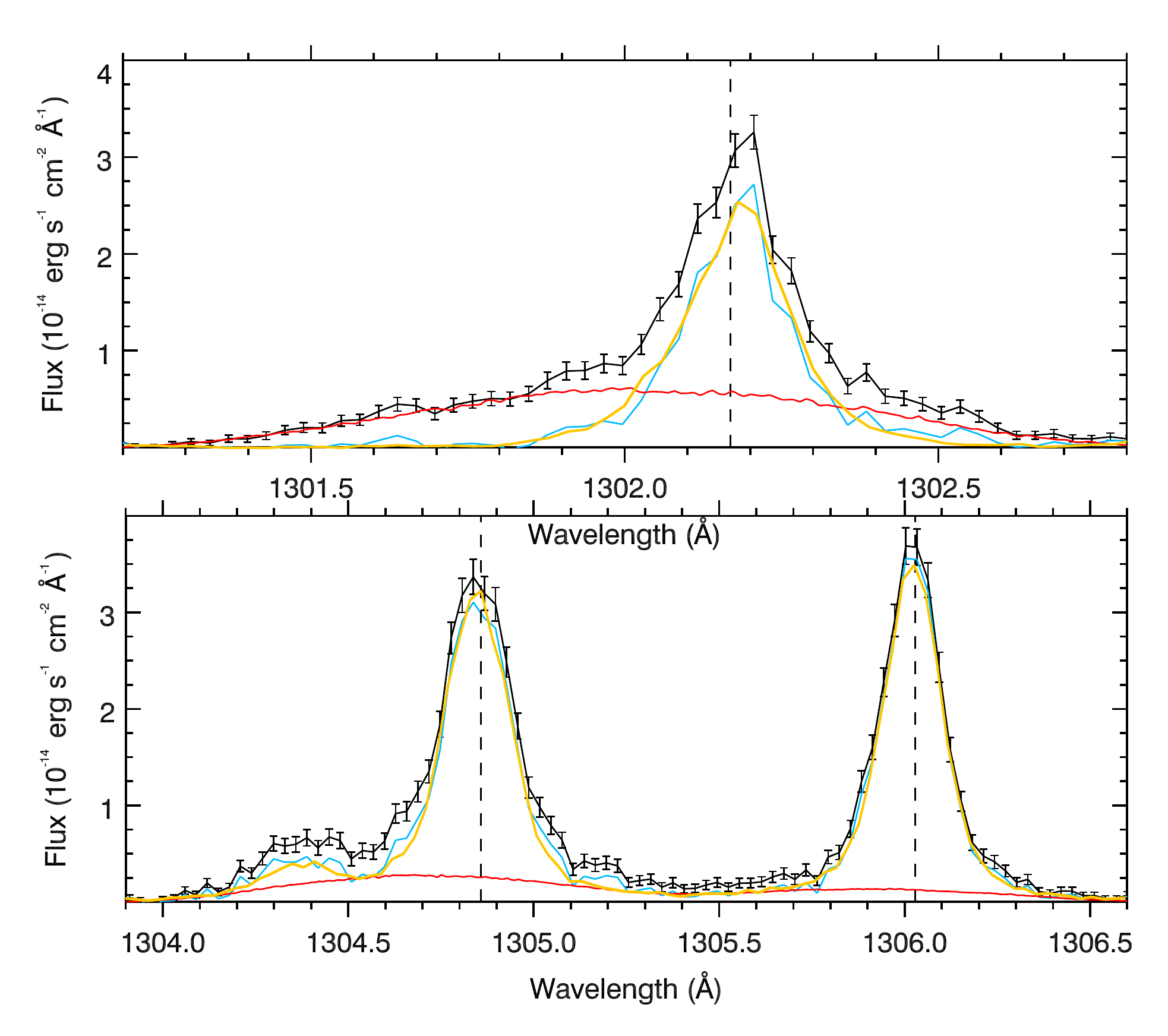}
\caption[]{Spectrum observed in Visit B$_\mathrm{fuv}$ in the region of the \ion{O}{i} triplet (average of the FP$_{4}$ over the visit, in black). Blue spectra show the stellar lines corrected for contamination using airglow templates (red profiles). They are in good agreement with the airglow-free spectra (average of the FP$_{1}$,$_{2}$,$_{3}$ over the visit, in orange). The stellar lines have been centered in the star rest frame (dashed black line).}
\label{fig:corr_OI_airglow}
\end{figure}
\end{center}

\begin{center}
\begin{figure}[tbh!]
\centering
\includegraphics[trim=2.5cm 6cm 2cm 12cm,clip=true,width=\columnwidth]{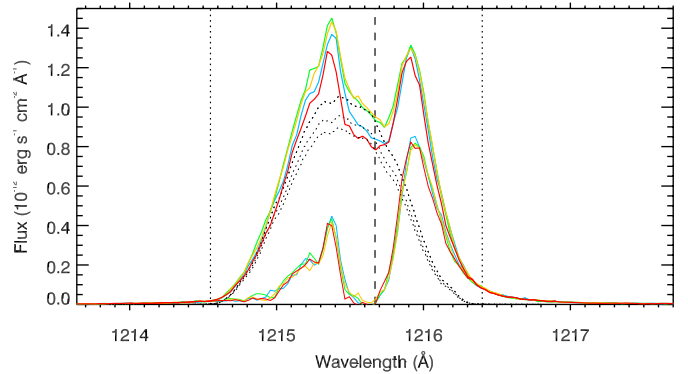}
\caption[]{Ly-$\alpha$ line of 55\,Cnc in Visit A$_\mathrm{fuv}$ (indicated in blue, green, orange, and red with increasing orbital phase), contaminated by geocoronal emission within the range delimited by vertical dotted lines. The stellar line is retrieved by subtracting observational airglow templates adjusted manually in each orbit (dotted black profiles). No absorption by the planet is observed in the extracted stellar line during the in-transit orbit (orange). The stellar line has been centered in the star rest frame (vertical dashed black line).}
\label{fig:corr_HI_airglow}
\end{figure}
\end{center}

Spectra are corrected for contamination by subtracting the airglow templates, adjusted in position and amplitude in each exposure. In the case of 55\,Cnc, the \ion{O}{i}$_\mathrm{air}$ $\lambda$1302 line is much broader than its stellar counterpart. This line could thus be fitted over the spectral range where it alone contributes to the observed flux (Fig.~\ref{fig:corr_OI_airglow}, upper panel). The excited \ion{O}{i} lines of 55\,Cnc, however, are blended together along with the \ion{Si}{ii} $\lambda$1304.37 line, preventing us from adjusting the airglow lines independently. Nonetheless, adjusting manually the templates allowed us to obtain spectra consistent with the airglow-free \ion{O}{i} stellar lines (Fig.~\ref{fig:corr_OI_airglow}, lower panel). The \ion{H}{i}$_\mathrm{air}$ line could be fitted to the observed spectra over the spectral range where the stellar flux is fully absorbed by the ISM, and there is a good agreement between the corrected Ly-$\alpha$ lines profiles of 55\,Cnc in each exposure (Fig.~\ref{fig:corr_HI_airglow}). \\

Our goal was to show that it is possible to extract the stellar lines to a good precision with such templates, even when spectra are strongly contaminated. At worst, when the templates cannot be fitted independently from the stellar spectum, the airglow line can be overcorrected and yield a lower limit on the stellar line flux. We did not perform a precise correction for the airglow because it is possible to analyze directly the airglow-free \ion{O}{i} stellar lines and the far wings of the stellar Lyman-$\alpha$ line (uncontaminated by the airglow outside of 1214.55 -- 1216.4\,\AA\,, Fig.~\ref{fig:corr_HI_airglow}), and because a preliminary analysis of the airglow-corrected Ly-$\alpha$ spectra did not reveal any transit variations (as expected from the non-detection of a hydrogen exosphere in previous HST/STIS Ly-$\alpha$ transit observations; \citealt{Ehrenreich2012}).\\


\section{Spectro-temporal analysis of 55\,Cnc FUV lines}
\label{sec:ttt} 

\subsection{Methods}
\label{sec:meth} 

We searched for flux variations in the brightest lines of the 55\,Cnc spectrum with a particular interest in variations that would be phased with the planet transit and repeatable over the different visits. This required us to define a reference for the quiescent stellar line that is unaffected by stellar variability or planetary occultation. Out-of-transit measurements are not necessarily representative of the quiescent stellar lines, therefore we compared all spectra together to identify those showing no significant variations within each visit. This was performed by searching for all features characterized by relative flux variations with S/N larger than 3, and extending over more than two COS resolution elements (about 0.13\,\AA\,, or 13 pixels). Once stable spectra were identified for each line, they were coadded into a master quiescent spectrum that was used to characterize the features detected in the variable spectra more accurately. We show in Fig.~\ref{fig:Lines_spec_grid} a grid of all quiescent and variable spectra and describe the analysis of each line in the following sections. We note that signatures of ISM absorption are detected in the stellar lines of \ion{O}{i} $\lambda$1302 and in the \ion{C}{ii} $\lambda$1334 doublet, but do not affect our search for time-variable flux variations over a visit. \\

When features that could correspond to absorption signatures were detected in a stellar line, we used a toy model to estimate the properties of the putative absorber. A Gaussian or Voigt profile representing the intrinsic stellar line is absorbed by an optically thin cloud of gas with occulting area $S_\mathrm{g}$ (normalized by the stellar surface), broadening parameter $b_\mathrm{g}$, column density $N_\mathrm{g}$, and radial velocity $v_\mathrm{g}$ in the star rest frame. The transmission spectrum of the cloud is defined as
\begin{align}
&T(\mathrm{v}) = 1  - \mathrm{S}_\mathrm{g}\,( 1 - e^{-\tau(v)} ), \\
&\mathrm{where}\,\tau(\mathrm{v}) = \frac{e^{2}\,f_\mathrm{osc}\,\lambda_{0}}{4\,\epsilon_{0}\,m_e \,c\,\sqrt{\pi}\,b_\mathrm{g}}\,N_\mathrm{g}\,e^{-\left(\frac{v-v_\mathrm{g}}{b_\mathrm{g}}\right)^2}  \nonumber.
\end{align}
We fit simultaneously the unocculted and absorbed stellar line model, convolved with the relevant COS LSF, to the quiescent and variable spectra. Possible origins for the absorbers are discussed in Sec.~\ref{sec:SPIs}. We note that the planetary disk of 55\,Cnc e yields a transit depth of about 350\,ppm, which is undetectable in our dataset.\\

\begin{figure*}
\centering
\begin{minipage}[h!]{\textwidth}
\includegraphics[trim=0cm 0cm 0cm 0cm,clip=true,width=\columnwidth]{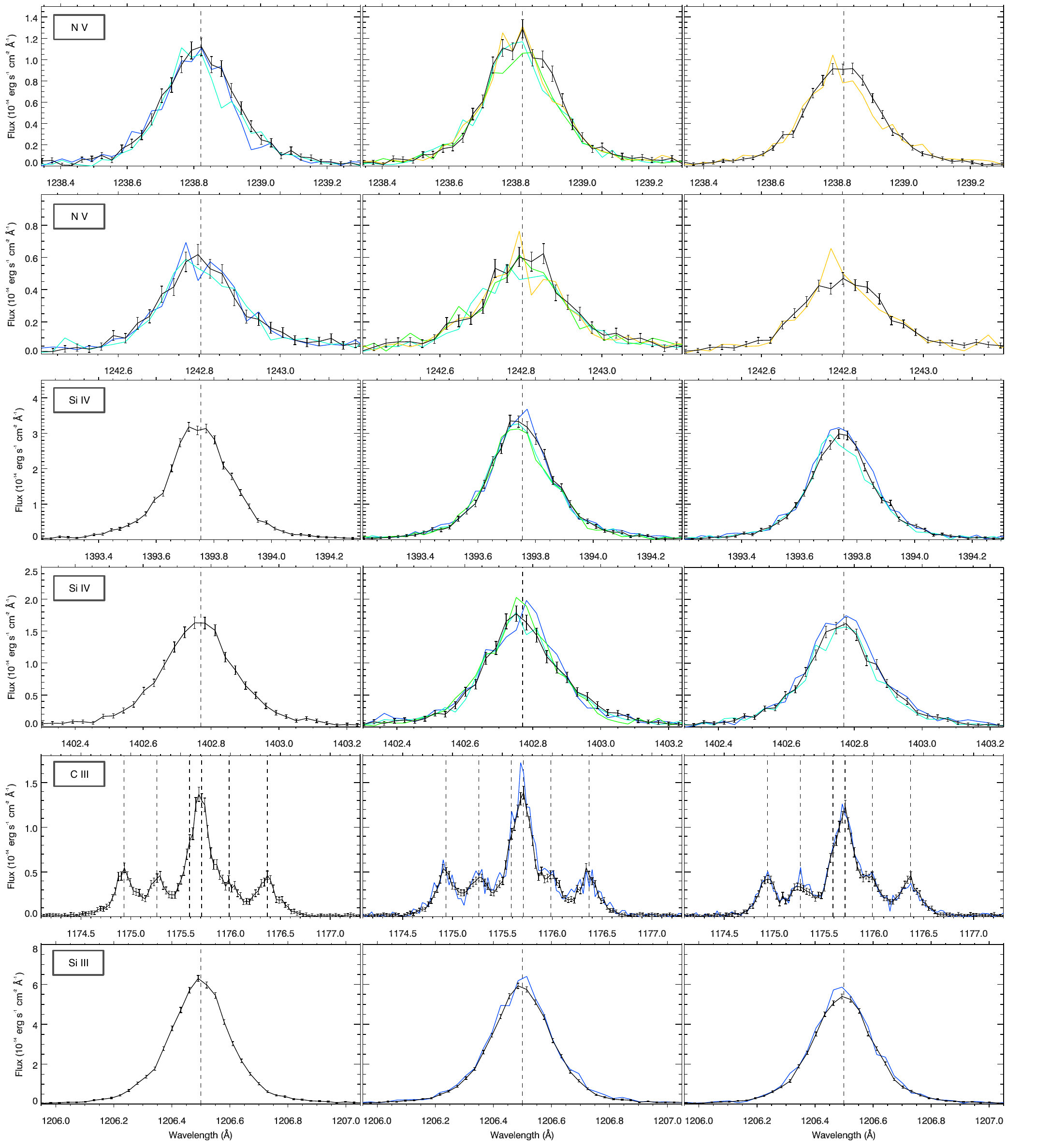}
\end{minipage}
\caption[]{Spectral profiles of 55\,Cnc FUV lines. Each row corresponds to a stellar line, ordered from top to bottom by decreasing formation temperature (Table~\ref{tab:studied_lines}). First to third columns correspond to visits A$_\mathrm{fuv}$, B$_\mathrm{fuv}$, and C$_\mathrm{fuv}$, respectively. Black profiles with error bars are the spectra taken as reference for the quiescent stellar lines, built by coadding stable spectra in each visit. Colored profiles correspond to spectra showing significant deviations from these references (exposures 0 to 4 in blue, cyan, green, orange, and red, respectively). All stellar lines have been corrected for their measured spectral shift and are thus centered on their rest wavelength (vertical dashed line).}
\label{fig:Lines_spec_grid}
\end{figure*}

\begin{figure*}\ContinuedFloat
\centering
\begin{minipage}[h!]{\textwidth}
\includegraphics[trim=0cm 0cm 0cm 0cm,clip=true,width=\columnwidth]{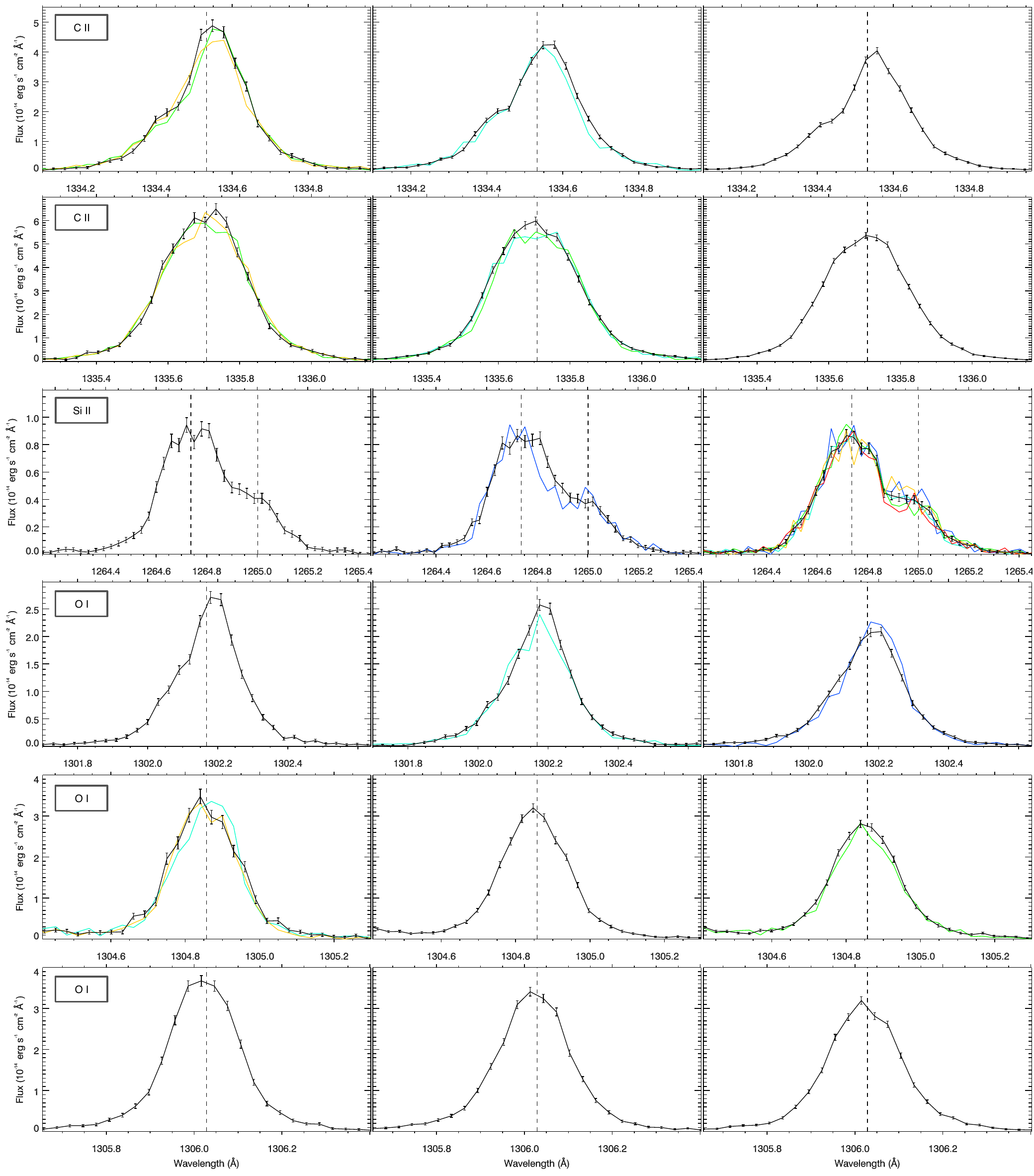}
\end{minipage}
\caption[]{Continued}
\label{fig:Lines_spec_grid}
\end{figure*}


\subsection{Repeatable features}
\label{sec:var_2017}

\subsubsection{Brightened stellar lines}
\label{sec:inc_em}

\begin{figure}[tbh!]
\centering
\includegraphics[trim=0cm 0cm 0cm 0cm,clip=true,width=\columnwidth]{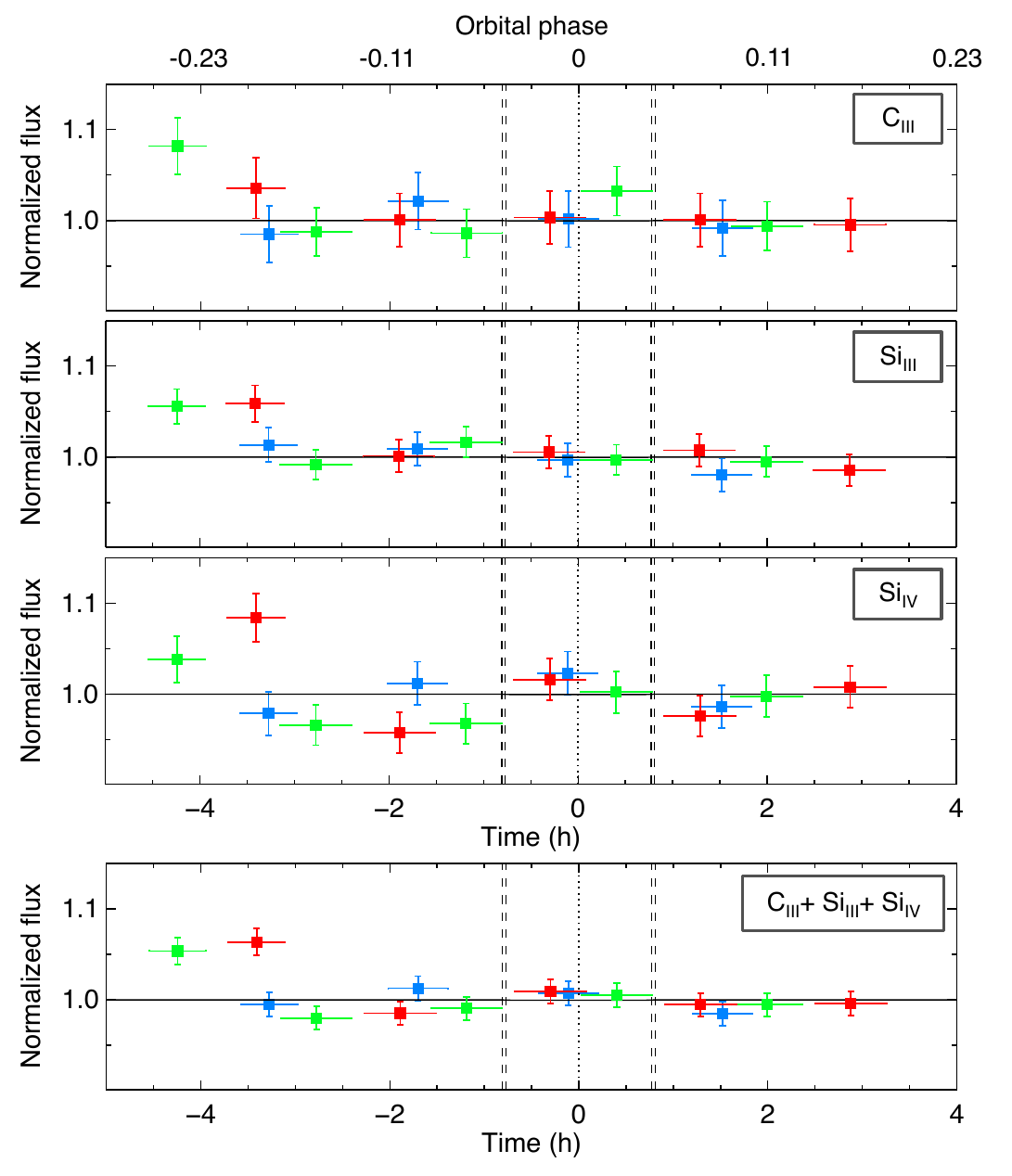}
\caption[]{Flux integrated over the full spectral range of the \ion{C}{iii}, \ion{Si}{iii}, and \ion{Si}{iv} lines (from top to bottom). The lowermost panel shows the flux cumulated over those lines. Fluxes are normalized using the quiescent spectra, averaged over all exposures in Visit A$_\mathrm{fuv}$ (blue), over exposures 1-4 in visits B$_\mathrm{fuv}$ (green) and C$_\mathrm{fuv}$ (red) for the \ion{C}{iii} and \ion{Si}{iii} lines, and over exposures 3-4 in visits B$_\mathrm{fuv}$ and C$_\mathrm{fuv}$ for the \ion{Si}{iv} line. Vertical dashed lines are the transit contacts of 55\,Cnc e. }
\label{fig:LC_CIII_SiIII_SiIV}
\end{figure}

The lines of \ion{C}{iii}, \ion{Si}{iii}, and \ion{Si}{iv} are stable over Visit A$_\mathrm{fuv}$ but are brighter in exposures obtained earlier than -3\,h (i.e., 3\,h before mid-transit) in both visits B$_\mathrm{fuv}$ and C$_\mathrm{fuv}$. The flux averaged over the three lines is about 6.0$\pm$1.5\% larger in each visit, compared to later orbits (see Fig.~\ref{fig:LC_CIII_SiIII_SiIV}). As can be seen in Fig.~\ref{fig:Lines_spec_grid} the \ion{Si}{iii} line is brighter over its entire profile, while the \ion{Si}{iv} doublet lines are brighter in their core. We cannot assess the behavior of individual lines in the \ion{C}{iii} multiplet because they are faint and blended. No variations are detected in the \ion{C}{iii} and \ion{Si}{iii} lines at later orbital phases.\\

\subsubsection{Absorption in the \ion{Si}{iv} doublet}
\label{sec:SiIVd}

\begin{figure}[tbh!]
\centering
\includegraphics[trim=0cm 0cm 0cm 0cm,clip=true,width=\columnwidth]{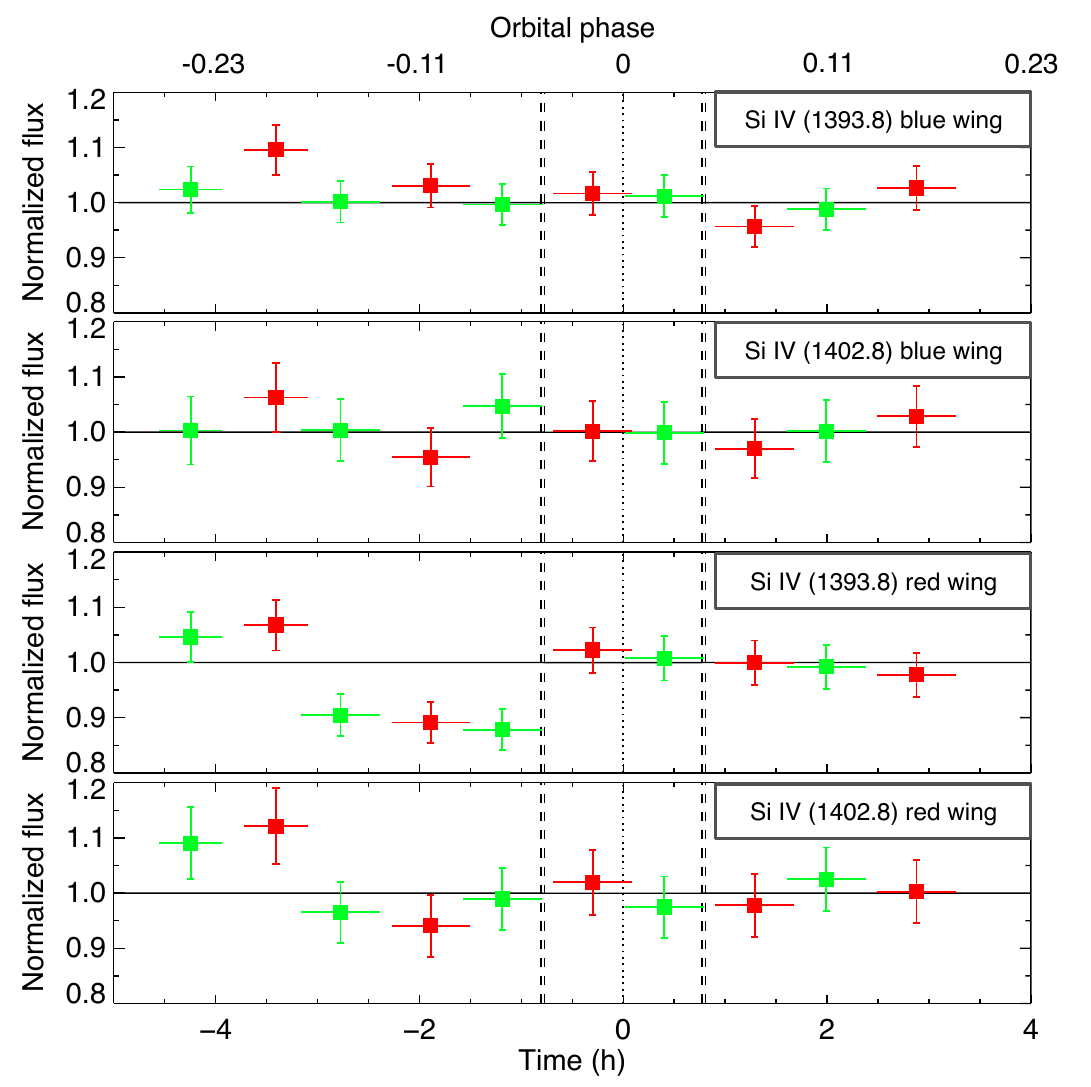}
\caption[]{Flux integrated in the blue wing (upper panels) and red wing (lower panels) of the \ion{Si}{iv} lines in Visit B$_\mathrm{fuv}$ (green) and Visit C$_\mathrm{fuv}$ (red). The absorption signature between -3 and -0.7h is diluted when measured over the entire red wing. Fluxes are normalized by the mean of exposures obtained after -0.7\,h (see text). }
\label{fig:LC_SiIV_V4V5}
\end{figure}

\begin{figure}[tbh!]
\centering
\includegraphics[trim=0cm 0cm 0cm 0cm,clip=true,width=\columnwidth]{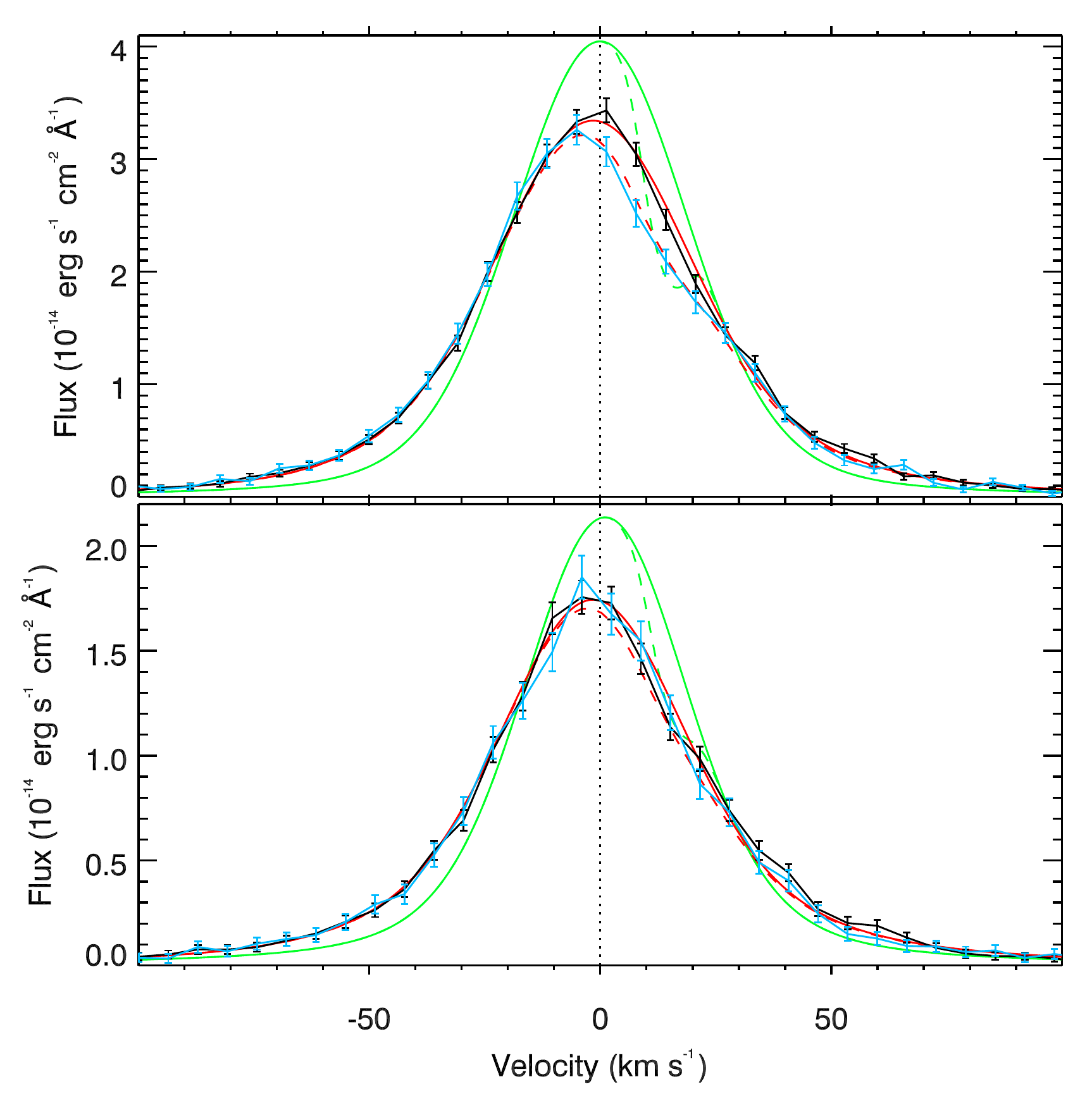}
\caption[]{\ion{Si}{iv} doublet averaged over visits B$_\mathrm{fuv}$ and C$_\mathrm{fuv}$, as a function of velocity relative to the line centers (vertical dotted lines). The upper panel shows the \ion{Si}{iv} $\lambda$1394 line. The lower panel shows the \ion{Si}{iv} $\lambda$1403 line. Pre-transit spectra (blue) show a flux decrease compared to the in- and post-transit spectra (black), consistent with absorption from an optically thin cloud of ionized silicon gas. The green (before convolution) and red (after convolution) profiles correspond to a best-fit model for the intrinsic stellar line (solid line) and the occulting cloud (dashed line).}
\label{fig:Sp_SiIV1_V4V5}
\end{figure}

Neither line of the \ion{Si}{iv} doublet show significant spectral variations in Visit A$_\mathrm{fuv}$ (Fig.~\ref{fig:Lines_spec_grid}). In visits B$_\mathrm{fuv}$ and C$_\mathrm{fuv}$ the spectra obtained during and after the optical transit are stable and were coadded to create quiescent spectra. Compared to these references, Fig.~\ref{fig:LC_SiIV_V4V5} shows that both lines of the doublet are brighter before -3\,h. The lines remain stable afterward, except for a significant flux decrease in the red wing of the \ion{Si}{iv} $\lambda$1394 line between -3\,h and -0.7\,h, which occurs within a similar spectral range in both visits (Fig.~\ref{fig:Lines_spec_grid}). We highlight this signature in Fig.~\ref{fig:Sp_SiIV1_V4V5} by comparing the mean of the three spectra showing absorption in visits B$_\mathrm{fuv}$ and C$_\mathrm{fuv}$ with the mean of the five remaining quiescent spectra. We measure an absorption of 12$\pm$2.4\% (5$\sigma$) between about -4 and 26\,km\,s$^{-1}$ in the \ion{Si}{iv} $\lambda$1394 line. There is no significant variation outside of this range (-0.2$\pm$2.2\%). No equivalent flux decrease is detected in the \ion{Si}{iv} $\lambda$1403 line.   \\

An optically thick cloud of matter would yield the same signature in both lines of the \ion{Si}{iv} doublet, which is not the case in visits B$_\mathrm{fuv}$ and C$_\mathrm{fuv}$. In contrast an optically thin gas cloud would absorb the \ion{Si}{iv} $\lambda$1394 line but remain undetected in the \ion{Si}{iv} $\lambda$1403 line because the ratio of the oscillator strengths of the lines is 2. The fainter line would be absorbed at the level of 6\% between -4 and 26\,km\,s$^{-1}$, which is consistent within 2$\sigma$ with the measured flux variation of -0.6$\pm$3.7\%. Using the model described in Sect.~\ref{sec:meth}, we found that the \ion{Si}{iv} doublet is well fitted with S$_\mathrm{g}\approxsup$0.6, $\log_\mathrm{10}$\,N$_\mathrm{g}\sim$12.5, b$_\mathrm{g}$ in 2.5 -- 7\,km\,s$^{-1}$, and v$_\mathrm{g}\sim$14\,km\,s$^{-1}$ (Fig.~\ref{fig:Sp_SiIV1_V4V5}). Although these results should be considered with caution given the uncertainties in the determination of the absorption signature, they provide first-order estimates on the properties of a putative gas cloud occulting the star (see Sect.~\ref{sec:meth}).\\

\subsubsection{Absorption in the \ion{Si}{ii} doublet}

\begin{figure}
\centering
\includegraphics[trim=2.5cm 6cm 2cm 12cm,clip=true,width=\columnwidth]{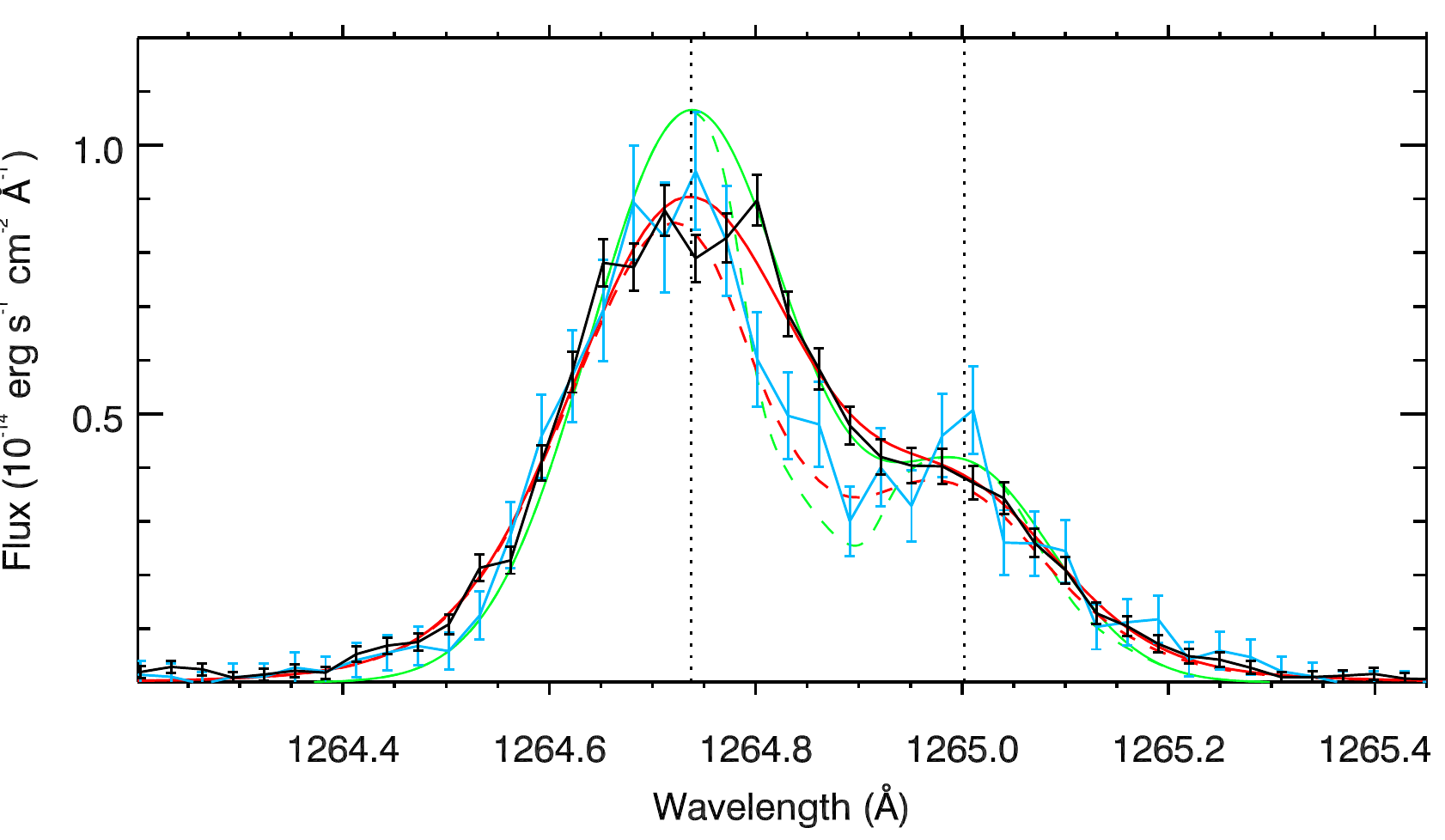}
\caption[]{\ion{Si}{ii} doublet in Visit B$_\mathrm{fuv}$ with the lines centered on their rest wavelengths (vertical dotted lines). The first exposure (blue) shows a flux decrease compared to the average of the other spectra in the visit (black), consistent with absorption from an optically thin cloud of silicon gas. The green (before convolution) and red (after convolution) profiles correspond to a best-fit model for the intrinsic stellar line (solid line) and the occulting cloud (dashed line).}
\label{fig:Sp_SiII}
\end{figure}

The two lines of the spin-orbit \ion{Si}{ii} doublet are blended and cannot be analyzed individually. No variations were found in Visit A$_\mathrm{fuv}$. No significant variations are observed in Visit B$_\mathrm{fuv}$ after -3\,h, but the first exposure shows a flux decrease of 27.4$\pm$6.0\% between 1264.79--1265.93\,\AA. As with the \ion{Si}{iv} doublet, this signature could arise from an optically thin gas cloud absorbing the \ion{Si}{ii} $\lambda$1264.7 line within about 12--45\,km\,s$^{-1}$. This scenario is consistent with the non-detection of the corresponding signature in the fainter \ion{Si}{ii} $\lambda$1265.0 line because the ratio of the oscillator strengths of the lines is about 10. We fitted our cloud model (Sect.~\ref{sec:meth}) to Visit B$_\mathrm{fuv}$ spectra and found that the detected signature is indeed well fitted to both doublet lines with S$_\mathrm{g}\approxsup$0.3, $\log_\mathrm{10}$\,N$_\mathrm{g}$ between about 12.6 and 14.4, b$_\mathrm{g}$ in 7 -- 19\,km\,s$^{-1}$, and v$_\mathrm{g}\sim$34\,km\,s$^{-1}$ (Fig.~\ref{fig:Sp_SiII}). The \ion{Si}{ii} doublet is mostly stable in Visit C$_\mathrm{fuv}$, but shows a strong variability in a localized spectral range ($\sim$ 1264.85 -- 1265.05\,\AA). It is not clear whether both lines are variable and in which exposures they deviate from the quiescent spectrum. We thus show in Fig.~\ref{fig:Lines_spec_grid} all spectra in Visit C$_\mathrm{fuv}$, as well as their mean that was taken as reference. We hypothesize that the signatures observed in the \ion{Si}{ii} and \ion{Si}{iv} doublets could be related (Sect.~\ref{sec:SPIs}).

\subsection{Persistent absorption in the \ion{N}{v} doublet}
\label{sec:var_NV}

\begin{figure}
\centering
\includegraphics[trim=0cm 0cm 0cm 0cm,clip=true,width=\columnwidth]{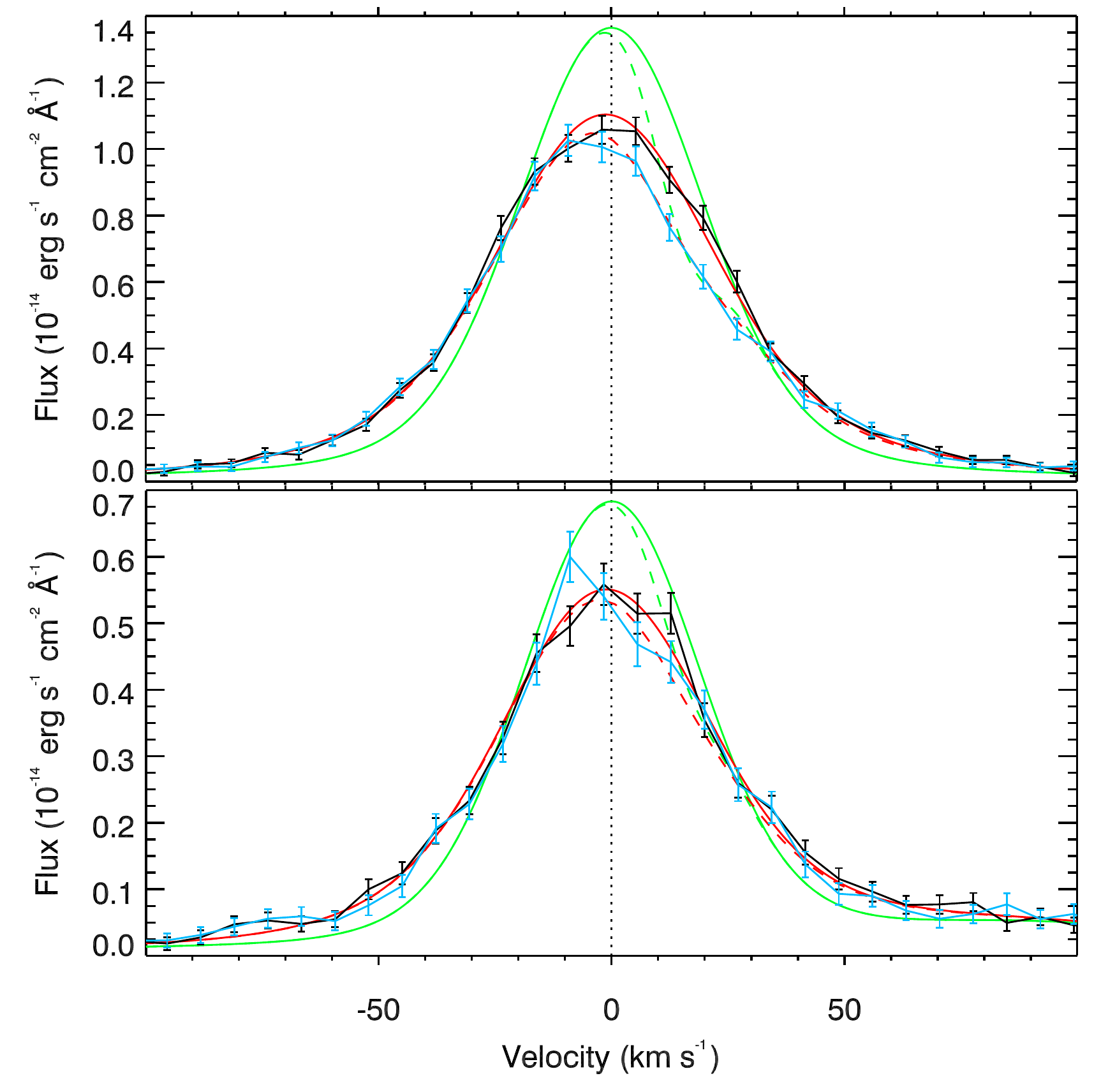}
\caption[]{\ion{N}{v} doublet averaged over all visits, as a function of velocity relative to the line centers (vertical dotted lines). The upper panel shows the \ion{N}{v} $\lambda$1238.8 line. The lower panel shows the \ion{N}{v} $\lambda$1242.8 line. Compared to the quiescent spectra (black), the absorbed spectra (blue) show a signature consistent with absorption from an optically thin cloud of ionized nitrogen gas. The green (before convolution) and red (after convolution) profiles correspond to a best-fit model for the intrinsic stellar line (solid line) and the occulting cloud (dashed line). The fit accounts for a weak stellar \ion{N}{i} line at 1243.18\,\AA, (90.5\,km\,s$^{-1}$ in the lower panel).}
\label{fig:Sp_NV_allvis}
\end{figure}

\begin{figure}
\centering
\includegraphics[trim=0cm 0cm 0cm 0cm,clip=true,width=\columnwidth]{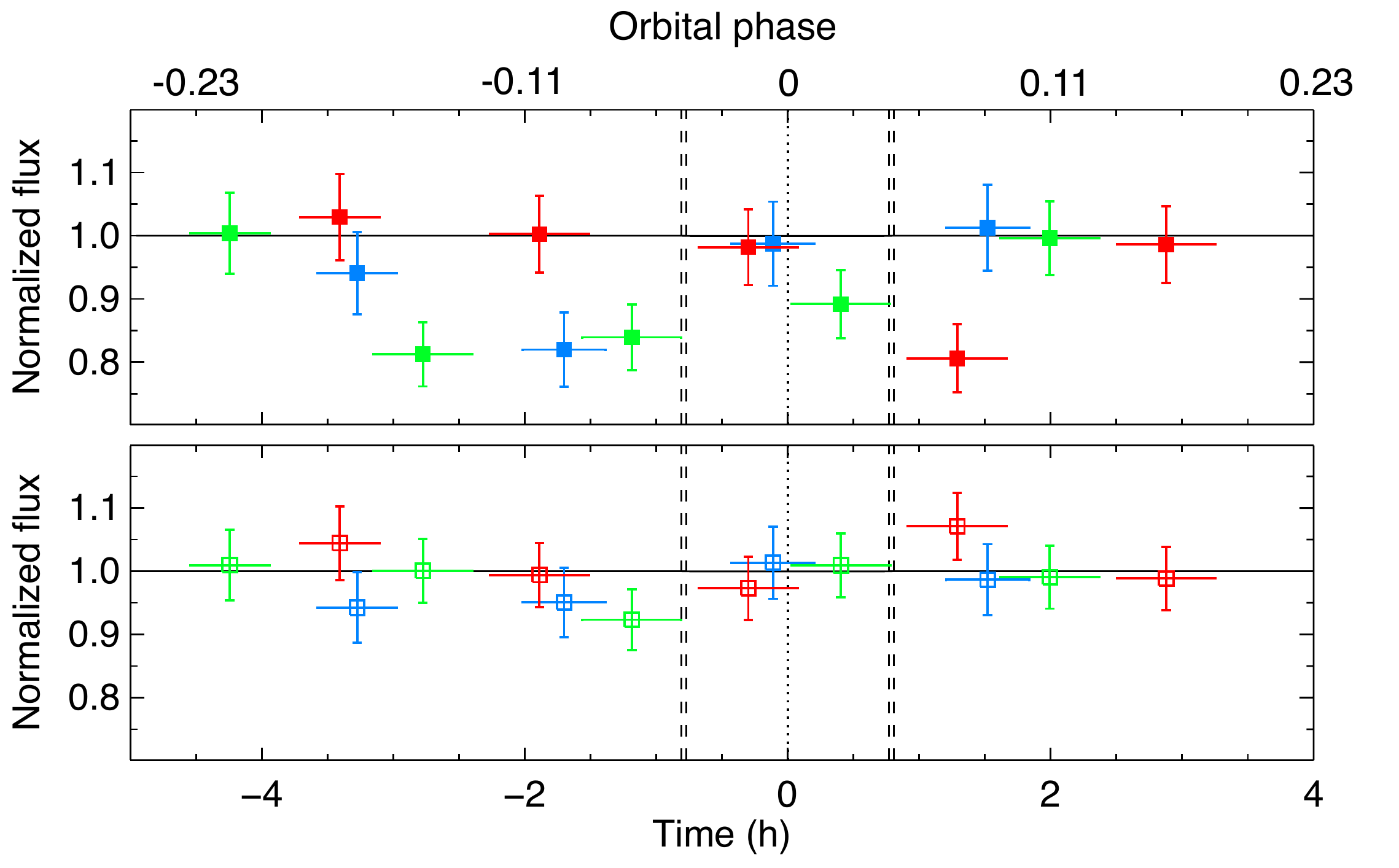}
\caption[]{\textit{Upper panel:} Flux in the \ion{N}{v} $\lambda$1238.8 line, integrated over the velocity range -3 and 31\,km\,s$^{-1}$, persistently absorbed in visits A$_\mathrm{fuv}$ (blue), B$_\mathrm{fuv}$ (green), and C$_\mathrm{fuv}$ (red). \textit{Lower panel}: Flux integrated over the complementary of the absorbed range within -100 to 100\,km\,s$^{-1}$. Fluxes are normalized using the stable reference spectra in each visit (see text) and plotted as a function of time relative to 55\,Cnc e transit.}
\label{fig:LC_NV1}
\end{figure}

The bright \ion{N}{v} $\lambda$1238.8 line shows strong, localized flux variations in each visit. We built quiescent spectra by averaging the following stable exposures: 2+3 in Visit A$_\mathrm{fuv}$, 0+4 in Visit B$_\mathrm{fuv}$, and 0+1+2+4 in Visit C$_\mathrm{fuv}$. Compared to these references, all other spectra show absorption signatures beyond the 3$\sigma$ level with depths ranging between about 15--25\%. Surprisingly these signatures occur at different orbital phases in each visit and their wavelength ranges vary within about 0 and 50\,km\,s$^{-1}$ in Visit A$_\mathrm{fuv}$, -25 and 40\,km\,s$^{-1}$ in Visit B$_\mathrm{fuv}$, and 0 and 35\,km\,s$^{-1}$ in Visit C$_\mathrm{fuv}$ (Fig.~\ref{fig:Lines_spec_grid}). This behavior strongly contrasts with the stable absorption signatures measured in the \ion{Si}{iv} doublet in visits B$_\mathrm{fuv}$ and C$_\mathrm{fuv}$ (Sect.~\ref{sec:SiIVd}). Nonetheless, all absorption signatures in the \ion{N}{v} $\lambda$1238.8 line overlap over a common wavelength range, at low velocity in the red wing. We compared the mean of the absorbed spectra with the mean of the quiescent spectra over all visits to quantify this persistent absorption (Fig.~\ref{fig:Sp_NV_allvis}), which occurs between about -3 and 31\,km\,s$^{-1}$ with a depth of 15$\pm$2.7\% (5.5$\sigma$). Interestingly, absorption consistently reaches a maximum depth of about 20\% in this spectral range, despite the different temporal evolution in each visit (Fig.~\ref{fig:LC_NV1}). There is no significant variation outside of the absorbed range (0.7$\pm$2.6\%, Fig.~\ref{fig:Sp_NV_allvis}). \\

The fainter \ion{N}{v} $\lambda$1242.8 line appears to be stable in all visits, as would be expected from an optically thin cloud of ionized nitrogen gas occulting the star. Indeed, the absorption signature detected in the \ion{N}{v} $\lambda$1238.8 line within -3 and 31\,km\,s$^{-1}$ would then yield an absorption of 7.5$\pm$1.4\% in the \ion{N}{v} $\lambda$1242.8 line (given the ratio of 2 between the lines oscillator strengths), which is consistent with the measured flux variation (6.0$\pm$4.3\%). As can be seen in Fig.~\ref{fig:Sp_NV_allvis} the spectra of both lines are well fitted by our cloud model (Sect.~\ref{sec:meth}) with S$_\mathrm{g}\approxsup$0.5, $\log_\mathrm{10}$\,N$_\mathrm{g}$ between about 13 and 13.5, b$_\mathrm{g}$ in 6.5 -- 12.5\,km\,s$^{-1}$, and v$_\mathrm{g}\sim$17\,km\,s$^{-1}$.  \\

\begin{center}
\begin{figure}[tbh!]
\centering
\includegraphics[trim=1.2cm 5cm 1cm 10cm,clip=true,width=\columnwidth]{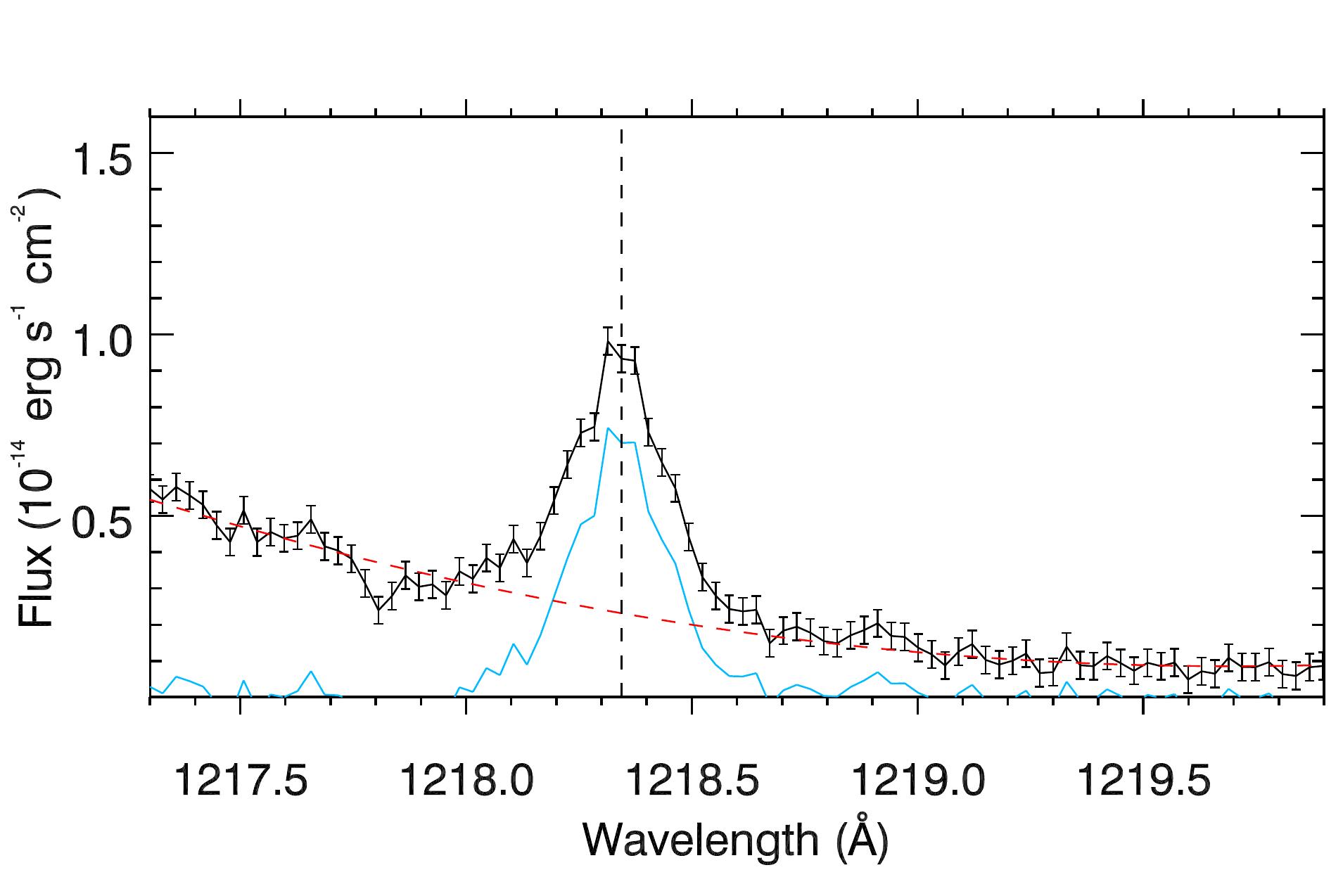}
\caption[]{\ion{O}{v} line in Visit A$_\mathrm{fuv}$ (blue). This line is superimposed on the red wing of the Ly-$\alpha$ line (black), which was corrected for using a polynomial function (dashed red line). The stellar line is centered on its rest wavelength (vertical dashed black line).}
\label{fig:OV_correction}
\end{figure}
\end{center}

\begin{figure}
\centering
\includegraphics[trim=0cm 0cm 0cm 0cm,clip=true,width=\columnwidth]{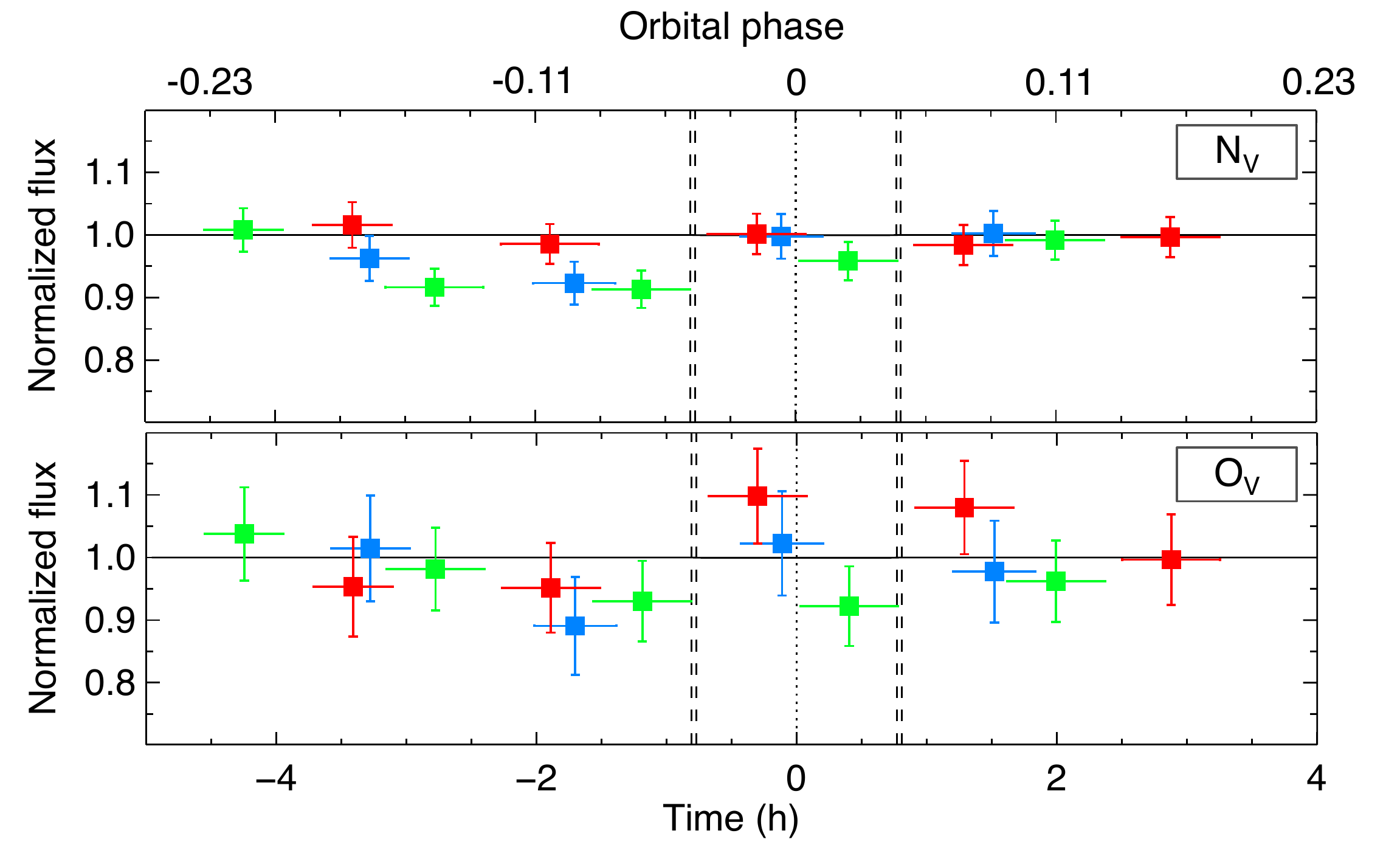}
\caption[]{Flux integrated over the entire \ion{N}{v} doublet (upper panel) and \ion{O}{v} line (lower panel). For the sake of comparison, fluxes are normalized to the same exposures as in Fig.~\ref{fig:LC_NV1} with  the same color code.}
\label{fig:LC_OV_NV}
\end{figure}

Finally, we searched the \ion{O}{v} line for variations similar to those detected in the \ion{N}{v} doublet. Both lines are indeed formed at similar temperatures in the transition region between the stellar chromosphere and corona (logT$\sim$5.2). In our observations the \ion{O}{v} line is superimposed with the red wing of the stellar Ly-$\alpha$ line, which was corrected for in each visit using a second order polynomial (Fig.~\ref{fig:OV_correction}). The \ion{O}{v} line is too faint to be analyzed spectrally, and we compared in Fig.~\ref{fig:LC_OV_NV} the flux integrated over the entire \ion{O}{v} and \ion{N}{v} lines. Unfortunately, the S/N of the integrated \ion{O}{v} line remains too low to assess the presence of variations correlated with that of the \ion{N}{v} doublet.\\

\subsection{Variability and absorption in the \ion{C}{ii} doublet}
\label{sec:CII_d}

The ISM absorption yields a clear signature in the blue wing of the ground-state \ion{C}{ii} $\lambda$1334.5 line and affects the excited \ion{C}{ii} $\lambda$1335.7 line (Fig.~\ref{fig:Lines_spec_grid}) to a lesser extent. These stable signatures do not affect the search for variability over transit timescales. While both lines of the \ion{C}{ii} doublet are extremely stable in Visit C$_\mathrm{fuv}$, we found significant but non-repeatable variations in visits A$_\mathrm{fuv}$ and B$_\mathrm{fuv}$. \\

\begin{figure}
\centering
\includegraphics[trim=0cm 0cm 0cm 0cm,clip=true,width=\columnwidth]{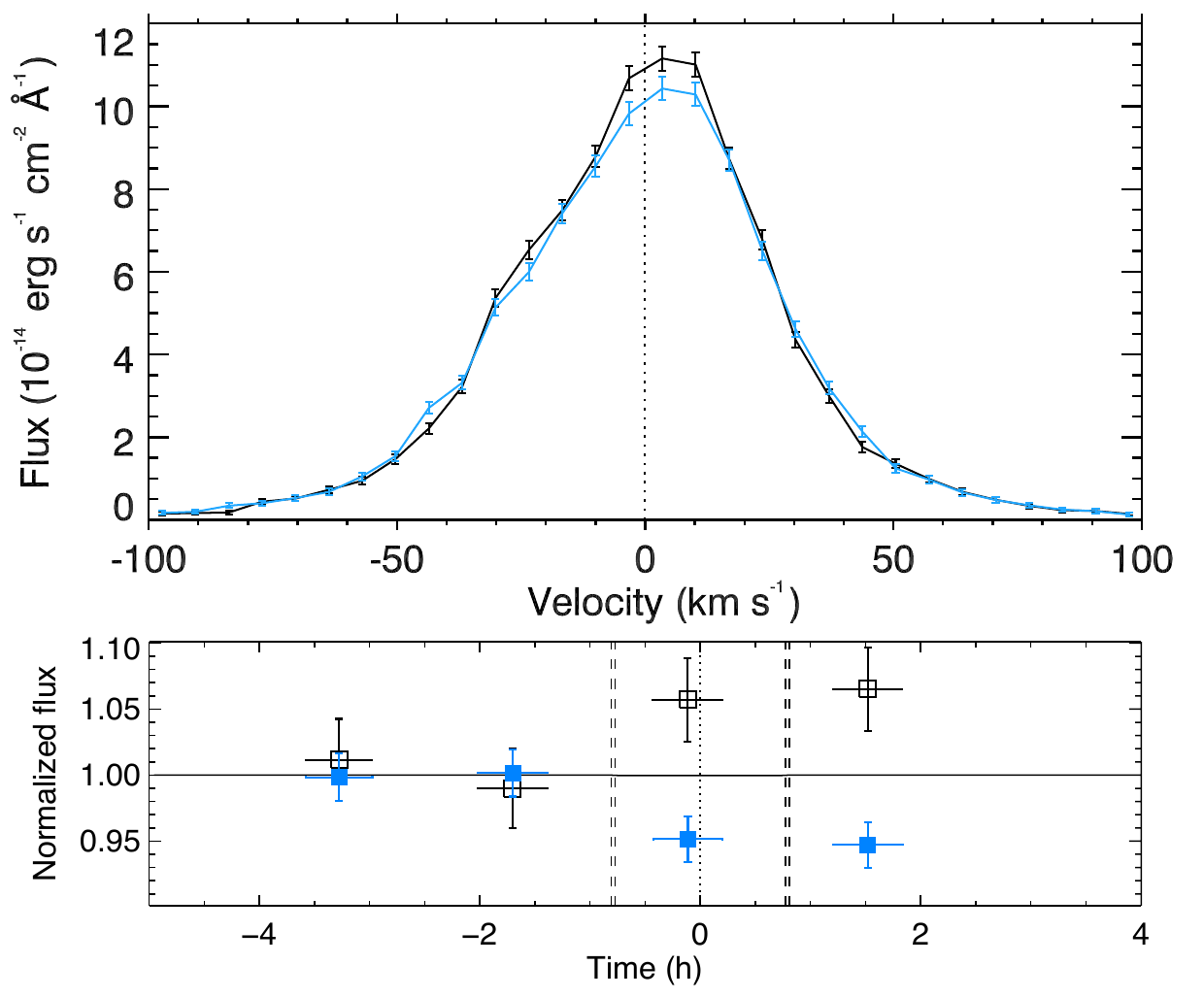}
\caption[]{\ion{C}{ii} doublet in Visit A$_\mathrm{fuv}$. \textit{Upper panel}: Coaddition of the $\lambda$1334.5 and $\lambda$1335.7 lines in velocity space. The comparison of the pre-transit mean spectrum (black) with the averaged in- and post-transit spectra (blue) reveals a flux decrease in the doublet core and a flux increase in its wings. \textit{Lower panel}: Flux integrated over the absorbed velocity range -29 to 27\,km\,s$^{-1}$ (filled squares) and over the complementary range within -100 to 100\,km\,s$^{-1}$ (empty squares). Fluxes are normalized by the pre-transit values and plotted as a function of time relative to 55\,Cnc e transit. }
\label{fig:CII_doublet_sp_lc_V2}
\end{figure}

In Visit A$_\mathrm{fuv}$ we averaged exposures 0 and 1 to build the quiescent spectrum for both lines of the doublet. Compared to these references, the \ion{C}{ii} $\lambda$1334.5 line decreases by 13.3$\pm$3.6\% between -31 and 2\,km\,s$^{-1}$ in exposure 2, and by 10.8$\pm$3.2\% between -4 and 27\,km\,s$^{-1}$ in exposure 3 (Fig.~\ref{fig:Lines_spec_grid}). No significant variations are measured in the \ion{C}{ii} $\lambda$1335.7 line, although marginal flux decreases are also found in its core for both exposures 2 and 3. Their average shows an absorption of 5.6$\pm$2.0\% between -29 and 16\,km\,s$^{-1}$, which is about two times lower than the signatures detected in the \ion{C}{ii} $\lambda$1334.5 line despite the doublet lines having similar oscillator strengths. Therefore, if these variations arise from a cloud of ionized carbon atoms occulting the star, most of these atoms would have to be in their fundamental state. Alternatively these variations could trace a change in the shape of the intrinsic stellar lines from exposures 0+1 to exposures 2+3 (Fig.~\ref{fig:Lines_spec_grid}), which we investigated by coadding the two lines of the doublet in velocity space (upper panel in Fig.~\ref{fig:CII_doublet_sp_lc_V2}). The flux decreases significantly in the core of the doublet between the two groups of exposures (4.9$\pm$1.4\% between -29 and 29\,km\,s$^{-1}$), while the wings of the doublet show a marginal increase by 6.0$\pm$2.2\%. It is noteworthy that both flux increase and decrease are present in exposures 2 and 3 individually and have similar amplitudes (Fig.~\ref{fig:CII_doublet_sp_lc_V2}). \\

\begin{figure}
\centering
\includegraphics[trim=0cm 0cm 0cm 0cm,clip=true,width=\columnwidth]{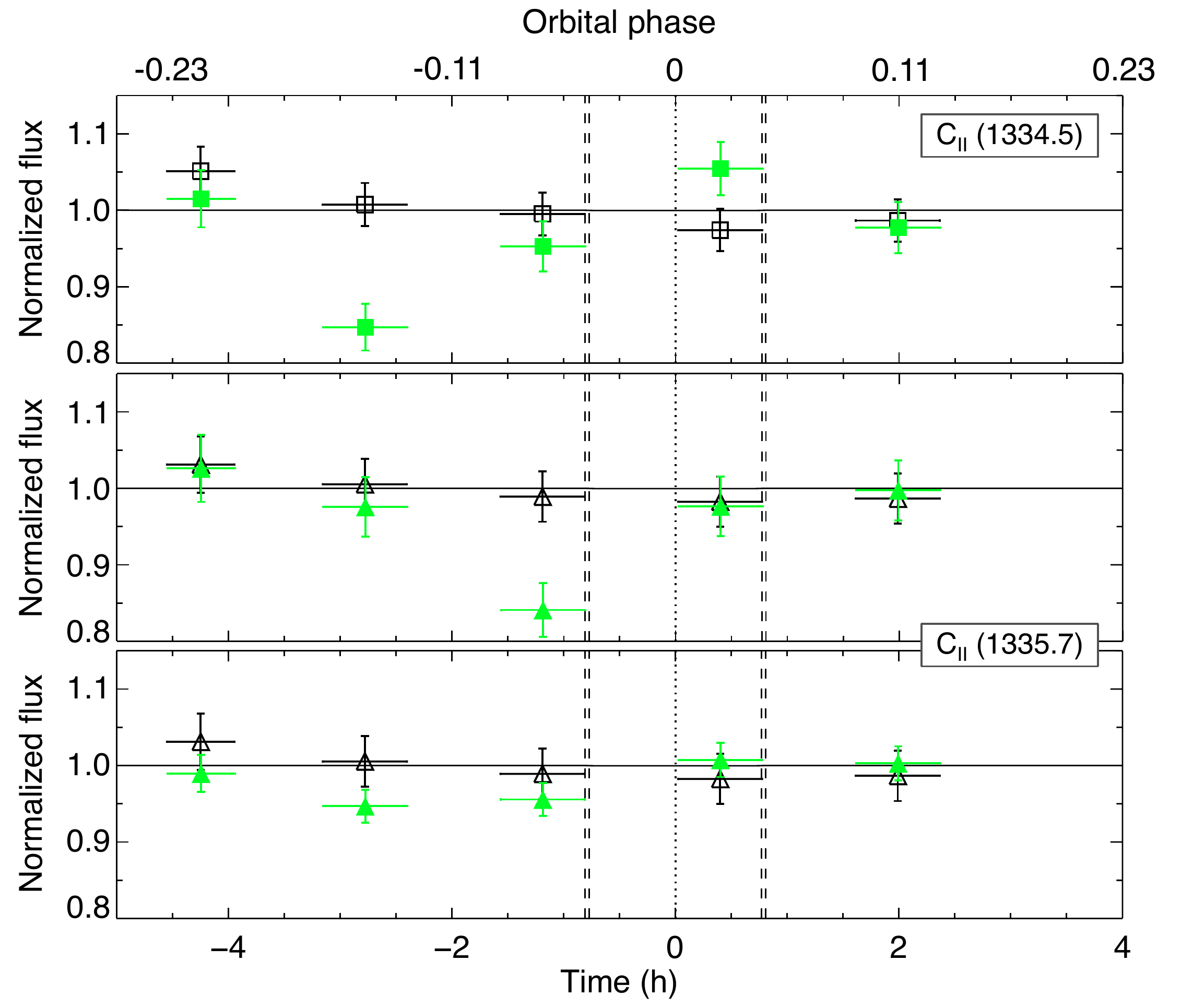}
\caption[]{Flux integrated over the stable (empty symbols) and absorbed spectral ranges (filled symbols) in the \ion{C}{ii} $\lambda$1334.5 (squares) and \ion{C}{ii} $\lambda$1335.7 (triangles) lines in Visit B$_\mathrm{fuv}$. Fluxes are normalized by their value in the quiescent spectra (see text). \textit{Upper panel}: The absorbed range extends from 5 to 41\,km\,s$^{-1}$. The stable range is its complementary within -100 to 100\,km\,s$^{-1}$. \textit{Lower panels}: The absorbed range extends from -53 to -22\,km\,s$^{-1}$ in the top panel and from -27 to 18\,km\,s$^{-1}$ in the bottom panel. The stable range is defined as [-100,-53]+[18,100]\,km\,s$^{-1}$. Absorption in the line core is diluted because the flux is integrated over a spectral range encompassing both exposures 1 and 2 signatures.}
\label{fig:CII_LC_V4}
\end{figure}

The behavior of the \ion{C}{ii} doublet is more complex in Visit B$_\mathrm{fuv}$. Exposure 1 shows a strong flux decrease in the ground-state \ion{C}{ii} $\lambda$1334.5 line, when compared to the average of the stable spectra in the other exposures (Fig.~\ref{fig:Lines_spec_grid}). This signature is located between 5 and 41\,km\,s$^{-1}$ with a depth of 15.3$\pm$3.1\% (5\,$\sigma$). In contrast the excited \ion{C}{ii} $\lambda$1335.7 line shows two different kinds of signatures, when taking the mean of the stable spectra in exposures 0, 3, and 4 as reference. Exposure 2 shows a flux decrease with similar depth (15.9$\pm$3.5\%) and width ($\sim$34\,km\,s$^{-1}$) as the signature detected in the ground-state line, but occurring one HST orbit later and shifted to negative velocities (-53 to -22\,km\,s$^{-1}$). Both exposures 1 and 2 show additional flux decreases in the core of the excited line (7.8$\pm$2.5\% from -27 to 5\,km\,s$^{-1}$, and 7.1$\pm$2.5\% from -13 to 18\,km\,s$^{-1}$, respectively). We show in Fig.~\ref{fig:CII_LC_V4} the flux integrated over the stable and absorbed spectral ranges of the \ion{C}{ii} $\lambda$1334.5 (upper panel) and the \ion{C}{ii} $\lambda$1335.7 (middle and lower panel) lines. The flux decreases in Visit B$_\mathrm{fuv}$ could correspond to absorption signatures, but the phase windows and spectral ranges of the signatures are very different between the two lines of the doublet and therefore we did not attempt to model them with our cloud toy model. \\

\subsection{Variability in the low-excitation \ion{O}{i} lines}
\label{sec:OItrip}

\begin{figure}[tbh]
\centering
\includegraphics[trim=0cm 0cm 0cm 0cm,clip=true,width=\columnwidth]{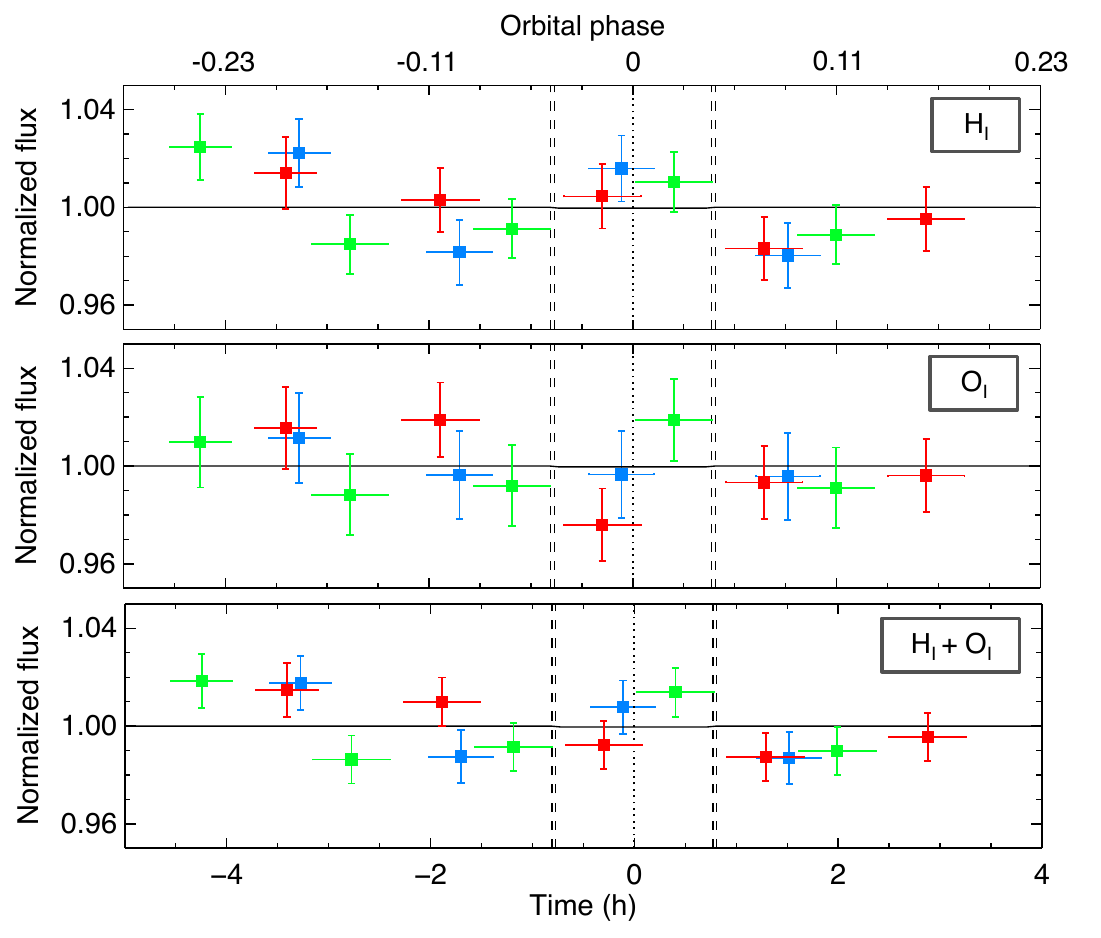}
\caption[]{Flux integrated over the airglow-free Lyman-$\alpha$ line wings (upper panel), the \ion{O}{i} triplet (middle panel), and the cumulated lines (lower panel) as a function of time relative to the transit of 55\,Cnc e. Visits A$_\mathrm{fuv}$, B$_\mathrm{fuv}$, and C$_\mathrm{fuv}$ are indicated in blue, green, and red.}
\label{fig:HI_OI}
\end{figure}

The core of the Lyman-$\alpha$ line is emitted at high temperatures (up to log\,T$\sim$4.5 in the transition region) while its wings probe deeper into the chromosphere as one observes farther from the line center (down to log\,T$\sim$3.9 beyond about 1\,\AA; \citealt{Vernazza1981}, \citealt{Roussel1982}, \citealt{Fontenla1991}). The airglow-free Lyman-$\alpha$ line wings of 55\,Cnc (beyond -1.1 and +0.7\,\AA\, from the line center, see Sect.~\ref{sec:air_corr}) should thus arise from similar altitudes in the chromosphere as the \ion{O}{i} triplet (log\,T$\sim$3.9). We show in Fig.~\ref{fig:HI_OI} the comparison between the flux integrated over the \ion{O}{i} triplet and over the Lyman-$\alpha$ line wings (integrated up to $\pm$1.9\,\AA\, from the line center, so as to avoid the \ion{O}{v} line). Neither line show signifiant deviations with respect to their mean over the visits. There is thus no evidence for a correlation between the lines in any of the visits beyond their common stability. In a second step, we performed a spectral analysis of each \ion{O}{i} triplet line. The ISM absorption observed in the blue wing of the ground-state \ion{O}{i}$\lambda$1302.2 line (Fig.~\ref{fig:Lines_spec_grid}) is consistent with the signature observed in the \ion{C}{ii} line. As expected, the excited \ion{O}{i} $\lambda$1304.8 and \ion{O}{i} $\lambda$1306.0 lines show no signatures of ISM absorption. \\

Visit A$_\mathrm{fuv}$ shows no significant deviations between any two exposures in the \ion{O}{i} $\lambda$1302.2 and \ion{O}{i} $\lambda$1306.0 lines. The \ion{O}{i} $\lambda$1304.8 line shows symmetric, consistent line profiles in exposures 0 and 2 but has a strongly redshifted peak in exposure 1, and a significantly lower flux beyond about $\pm$20\,km\,s$^{-1}$ in exposures 1 and 3. Visit B$_\mathrm{fuv}$ shows no significant deviations between any two exposures in the \ion{O}{i} $\lambda$1304.8 and \ion{O}{i} $\lambda$1306.0 lines. The \ion{O}{i} $\lambda$1302.2 line is stable except for exposure 1, which shows a flux decrease of 12.7$\pm$4.0\% between -11 and 21\,km\,s$^{-1}$. In Visit C$_\mathrm{fuv}$ only the \ion{O}{i} $\lambda$1306.0 line is stable in all exposures. The \ion{O}{i} $\lambda$1302.2 line shows a redshifted peak and significantly lower flux in the blue wing in the first exposure and is stable afterward. The \ion{O}{i} $\lambda$1304.8 line is stable, except for a deep absorption signature in exposure 2 (8.0$\pm$2.7\% between -38 and 27\,km\,s$^{-1}$). These asymmetries and absorptions in the \ion{O}{i} triplet can be seen in Fig.~\ref{fig:Lines_spec_grid}. They occur in the core of the lines and in their wings, and display no apparent correlation between the lines or the visits. Interestingly, though, the \ion{O}{i} $\lambda$1306.0 line shows no variability in any of the visits, even though it is  the brightest line in the triplet and arises from the highest excitation level. This suggests that variations in the \ion{O}{i} $\lambda$1302.2 and \ion{O}{i} $\lambda$1304.8 lines arise from physical changes in a population of neutral oxygen atoms in the ground-state or lower excitation level.\\


\subsection{Summary of main features}
\label{sec:summary} 

We summarize here the variations measured in 55\,Cnc FUV lines. First, we note three features common to all visits: 
\begin{itemize}
\item \textit{Variability in the \ion{O}{i} triplet:} This variability is limited to the ground-state and lower-excitation \ion{O}{i} lines, but otherwise show no obvious pattern. The lines are either distorted or show absorption in different spectral regions, and these changes occur in non-successive exposures and at different orbital phases in each visit.
\item \textit{Absorption signatures:} All signatures consistent with absorption of the quiescent stellar lines (for \ion{Si}{ii}, \ion{Si}{iv}, \ion{C}{ii}, \ion{N}{v}) occur in successive or single exposures. Except for the excited \ion{C}{ii} line in Visit A$_\mathrm{fuv}$, all signatures are located in the core or the red wing of the stellar lines and are better explained by absorption from an optically thin gas cloud rather than an opaque disk. Except for the \ion{N}{v} line in Visit C$_\mathrm{fuv}$, all signatures occur before or during the transit of 55\,Cnc e.
\item \textit{The \ion{N}{v} doublet:} This is the only line that shows signatures consistent with absorption in the three visits. 
\end{itemize}

Visit A$_\mathrm{fuv}$ only shows variability in three lines (\ion{O}{i}, \ion{N}{v}, \ion{C}{ii}) and thus contrasts with Visit B$_\mathrm{fuv}$ and C$_\mathrm{fuv}$, which show variations in most lines. These two visits further display some intriguing similarities, in particular a brighter emission before about -3\,h in a group of lines formed at similar chromospheric temperature (\ion{C}{iii}, \ion{Si}{iii}, \ion{Si}{iv}; log\,T$\sim$4.7-4.9; Table~\ref{tab:studied_lines}), followed by a flux decrease in the \ion{Si}{iv} doublet consistent with absorption from a silicon cloud between about -3 and -0.7\,h. Both features could be linked to the absorption in the \ion{Si}{ii} doublet occurring before -4\,h in Visit B$_\mathrm{fuv}$, as this phase range is not covered in Visit C$_\mathrm{fuv}$. \\

Finally, we note the peculiar behavior of the \ion{C}{ii} doublet. Visit A$_\mathrm{fuv}$ shows an overall change in the line shape over time that is not found in any other line and is not consistent with absorption alone. Visit B$_\mathrm{fuv}$ shows absorption in the core of both doublet lines and a deep absorption signature shifting from the red wing of the ground-state line to the blue wing of the excited line in two successive exposures. This is the only case of absorption at high negative velocities in our observations. Visit C$_\mathrm{fuv}$ shows no detectable variations.\\


\section{Interpreting the variations in 55\,Cnc FUV lines}
\label{sec:SPIs} 

\subsection{An unlikely planetary origin}
\label{sec:unlik_plorig}

Given its high temperature ($\sim$2000\,K, \citealt{Demory2016b}), 55\,Cnc~e could be partially covered by magma oceans (\citealt{Gelman2011}; \citealt{Elkins2012}) while tidal interactions with the star and companion planets could induce significant volcanism (\citealt{Jackson2008}; \citealt{Barnes2010}). Vapors from the magma (e.g., \citealt{Schaefer2009}) sputtering from the partially rocky surface (e.g., \citealt{Mura2011}) and volcanic plumes would inject dust and metals into an envelope surrounding the planet, which could provide a source for ion clouds occulting the star in the FUV. Although there is no evidence for a hydrogen-rich envelope (\citealt{Ehrenreich2012}, this work) that could expand and carry heavy species out of the deep gravitational well of the super-Earth (\citealt{VM2013}, \citealt{PerezChiang2013}; \citealt{Rappaport2014}), other scenarios could lead to the formation of an occulting exospheric cloud. A heavyweight atmosphere might be subjected to hydrodynamical escape given the strong XUV irradiation of the planet (see the case of a CO$_\mathrm{2}$-dominated atmosphere in \citealt{tian2009}). An exosphere could also result from a sputtering process that is similar to but stronger than that of Mercury (\citealt{RiddenHarper2016}). Interactions between 55\,Cnc e and the stellar magnetosphere, such as those that exist between Io and Jupiter (e.g., \citealt{Brown1994}), could even lead to significant escape of ionized species from the super-Earth. However, the observed absorption signatures would require a very extended, dense exosphere with an occulting area larger than $\sim$30\% of the stellar disk and column density in between 10$^{\mathrm{12.5}}$ and 10$^{\mathrm{14.5}}$\,cm$^{-2}$. Therefore, even though 55\,Cnc e likely formed in an environment rich in silicates and possibly rich in carbon (\citealt{Delgado2010}, but see \citealt{Teske2013}), any mechanism sufficiently efficient to sustain such an exosphere likely depleted the crust and mineral atmosphere in volatile elements (such as H, C, or N) over time. The super-Earth would now be dominated by heavier species (such as Na, Si, and Mg; see also \citealt{Dorn2017}), which makes it unlikely that it is the source for material yielding the absorption signatures detected in the lines of ionized carbon and nitrogen.\\

Secondly, we would expect material escaping from 55\,Cnc e to co-move with the planet, which would shift the spectral range of its absorption signature to increasing radial velocities during our visits. Because of the large orbital velocity of the planet (Sect.~\ref{sec:obs_desc}), absorption signatures would shift noticeably between successive exposures and even be blurred over a wide spectral range during a single HST exposure. This contrasts with the stability of the detected signals, which mostly remain within similar spectral ranges even over two HST exposures (eg Sect.~\ref{sec:SiIVd}). From a dynamical/spectral point of view, it is thus unlikely that the occulting material originates from the planet. \\

Finally, we investigated the photoionization lifetime of species of interest under the high-energy radiation of 55\,Cnc. Most of the stellar EUV spectrum is not observable from Earth because of ISM absorption, and we thus reconstructed the entire XUV spectrum up to 1600\,\AA\, using the differential emission measure (DEM) retrieval technique described in \citet{Louden2017}. This reconstruction is based on the direct measurements of the stellar FUV lines presented in this paper and on archival X-ray and Ly-$\alpha$ observations (\citealt{Ehrenreich2012}). The lifetimes, calculated with photoionization cross-section formulae from \citet{Verner1996}, are given in Table~\ref{tab:photoion_lt}. Given the velocity range of the measured absorption signatures and their short-term temporal variability, it does not appear possible for silicon and nitrogen atoms escaped from the planet to remain within its vicinity long enough to be ionized into Si$^\mathrm{3+}$ and N$^\mathrm{4+}$ ions (about 4 and 26\,hours, respectively). Furthermore, even though neutral carbon atoms would be ionized in a few minutes it would take only about 20\,min to lose their second electron, which is short compared to the $\sim$2\,hours timespan over which \ion{C}{ii} absorption is observed in Vist B$_\mathrm{fuv}$ (Sect.~\ref{sec:CII_d}). This further shows that the occulting material is unlikely to originate from the planet.  \\

\begin{table}[h!]
\caption[]{Photoionization lifetimes under 55\,Cnc irradiation}
\centering
\begin{threeparttable}
\begin{tabular}{lcc}
\hline
\hline
Species & Lifetime at semimajor axis (min)  \\
\hline
C$^\mathrm{0}$ & 2.6   \\
C$^\mathrm{+}$ & 23   \\
Si$^\mathrm{0}$ & 0.36   \\
Si$^\mathrm{+}$ & 91  \\
Si$^\mathrm{2+}$ & 240  \\
Si$^\mathrm{3+}$ & 640  \\
N$^\mathrm{0}$ & 4.7   \\
N$^\mathrm{+}$ & 19   \\
N$^\mathrm{2+}$ & 74 \\
N$^\mathrm{3+}$ & 1562  \\
N$^\mathrm{4+}$ & 5003  \\
    \hline
  \end{tabular}
  \begin{tablenotes}[para,flushleft]
  Notes: Calculations used a$_\mathrm{pl}$ = 0.01544\,au for the semimajor axis of 55\,Cnc e (\citealt{Demory2016b}). 
  \end{tablenotes}
  \end{threeparttable}
\label{tab:photoion_lt}
\end{table}


\subsection{Could all variations be intrinsic to the star ?}
\label{sec:unl_storig}

Intrinsic variations in the chromospheric structure of 55\,Cnc could explain the change in the shape of the \ion{C}{ii} doublet in Visit A$_\mathrm{fuv}$ and in the \ion{O}{i} lines in all visits (Sect.~\ref{sec:summary}). However, this origin is unlikely for the signatures in other lines, which repeat between Visits B$_\mathrm{fuv}$ and C$_\mathrm{fuv}$. As shown in Sect.~\ref{sec:summary}, most signatures correspond to absorption of the quiescent stellar lines with typical depth of 10-20\%, localized at low/positive radial velocities, and occurring before or during the transit of 55\,Cnc\,e. A full coverage of the planet orbit is needed to confirm that these variations are indeed phased with the transit, but their behavior so far appears to be consistent with occultations of the star by clouds of gas located above the chromosphere, rather than random variations in the intrinsic stellar lines. \\

We have shown in the previous section that the occulting material is unlikely to originate from the planet. A star can occult its own chromospheric emission through filaments, which are made of partially ionized plasma and are suspended above the chromosphere by magnetic field line loops. Solar filaments, for example, can be a hundred times cooler and denser than the coronal material in which they are immersed (e.g., \citealt{Parenti2014}). However, filaments take a few days to form, then remain mostly stable before disappearing through a slow decay or violent eruptions that eject plasma with upward speeds of more than 100\,km\,s$^{-1}$ (e.g., \citealt{Wang2010}). This behavior is not consistent with the observed absorption signatures, which disappear after a few hours and trace gas falling toward 55\,Cnc at velocities on the order of 10\,km\,s$^{-1}$. Thus, we do not know of any mechanism intrinsic to this star that could explain the variations observed in most of its chromospheric lines.\\


\subsection{Cool coronal rain induced by 55\,Cnc e}
\label{sec:cool_rain}

In summary, the gas clouds occulting 55\,Cnc chromosphere likely originate from the star, yet their formation appear to be linked with the orbital motion of 55\,Cnc e. Thus, we propose a scenario in which the super-Earth influences the structure of the stellar corona without the need for the planet to be the source for the occulting material. The detected signatures would trace a form of coronal rain, a process that is observed in the Sun (e.g., \citealt{Kawaguchi1970}, \citealt{Leroy1972}, \citealt{Kohutova2016}) and possibly in other stars (e.g., \citealt{Ayres2010,Ayres2015}) over timescales of a few hours (\citealt{Antolin2010}; \citealt{Antolin2012}). Thermal instability, occurring for example after a flare, leads to catastrophic cooling of gas suspended high in magnetic loops (\citealt{Muller2003,Muller2004,Muller2005}). The cool plasma then forms dense downflows sliding along the sides of the magnetic loops at velocity from few up to $\sim$200\,km\,s$^{-1}$ (e.g., \citealt{Schrijver2001}). Coronal rain is usually observed in emission in cool optical (such as H$\alpha$ or \ion{Ca}{ii}) or hot UV (such as \ion{Si}{iv}, \ion{C}{iv}; \citealt{Ayres2015}) chromospheric lines, but also in absorption with temperatures down to about 10$^{4}$\,K (\citealt{Schrijver2001}). \\

\subsubsection{Origin of the coronal rain}

The semimajor axis of 55\,Cnc e is only 3.5 stellar radii. Thus, we hypothesize that the motion of the planet at the fringes of the stellar corona leads to its destabilization and the formation of short-lived coronal precipitations phased with the orbital motion. This destabilization could arise from a variety of mechanisms: energy could be released into the corona because of reconnections between planetary and stellar magnetic fields (e.g., \citealt{Lanza2009,Lanza2012}); material with no signatures in our FUV observations could escape from 55\,Cnc e and be funneled into the corona along stellar magnetic field lines connecting with the planet (e.g., \citealt{Adams2011}, \citealt{Shkolnik2008}, \citealt{Pillitteri2014}); alfv{\'e}nic perturbations generated by 55\,Cnc e could propagate through the magnetized stellar wind down to the stellar chromosphere (e.g., \citealt{Ip2004}, \citealt{Preusse2006}, \citealt{Kopp2011}, \citealt{Strugarek2015}); planetary or stellar flows could be disrupted by the formation of a shock ahead of the planet and accrete onto the corona (e.g., \citealt{Matsakos2015}). There are thus various ways in which 55\,Cnc e could be responsible for the onset of coronal rain. We note that the observed chromospheric signatures cannot arise directly from a bow shock. While the hot, dense, and ionized material formed in a bow shock has been proposed as the source for early-ingress absorption in a few exoplanet systems (e.g., \citealt{Vidotto2010}, \citealt{Llama2013}, \citealt{Cauley2015}), the occultation of 55\,Cnc chromosphere occurs as early as -3\,h (about -60$^{\circ}$) before mid-transit. This is likely too far from the planet for a bow shock to develop, as it would require an extremely strong planetary magnetic moment or a slow and low-density stellar wind (e.g., \citealt{griessmeir2004}; \citealt{See2014}). We would further expect bow-shock material to co-move with the planet, leading to strong radial velocity shifts of its absorption signature over the duration of our observations. \\

\subsubsection{Signatures of the coronal rain}

Downflows associated with solar coronal rain can be hot enough to yield brightened, broadened, and redshifted emission line profiles (e.g., \citealt{Lacatus2017}). Issues with COS spectral calibration prevented us from searching for redshifts in 55\,Cnc emission lines (Sec.~\ref{sec:flux_w_cal}), but Visits B$_\mathrm{fuv}$ and C$_\mathrm{fuv}$ show an overall brightening of high-temperature chromospheric lines (Sect.~\ref{sec:inc_em}), which is followed by deep and relatively narrow absorption signatures in a larger set of chromospheric lines (Sect.~\ref{sec:summary}). These latter signatures are well explained by occulting gas clouds with areas larger than $\sim$30\% of the star, column densities in between 10$^{\mathrm{12.5}}$ and 10$^{\mathrm{14.5}}$\,cm$^{-2}$, and temperatures in between 10$^{4}$ and 5$\times$10$^{5}$\,K (Sect.~\ref{sec:ttt}). This suggests a scenario in which the hot, destabilized coronal material of 55\,Cnc cools down to chromospheric temperatures, thus becoming visible in emission up to about -3\,h, before falling down to lower altitude and further cooling down, thus becoming visible through absorption at low/positive velocities. The depth of the measured absorption signatures would require dense downflows spread over a much larger area of the star than in the case of solar coronal rain. The mechanism inducing this ``cool'' coronal rain in 55\,Cnc might thus wrench hot plasma from coronal magnetic loops at higher altitudes over large regions, before it cools down and falls back toward the chromosphere. \\

The phasing of the observed signatures before the planet transit suggests that coronal rain is triggered by the planet ahead of its motion (e.g., if set by an accretion stream or a bent magnetic connection), appears to be earlier than the transit because of the propagation time of alfv{\'e}nic waves from the planet to the star, or is visible at specific orbital phases because of geometry and projection effects despite being generated along the entire orbit of 55\,Cnc~e. Alternatively the inhomogeneous structure of the stellar magnetosphere and plasma distribution could allow for coronal rain in specific regions of the star. \\

\subsubsection{Variability of the coronal rain}

The inhomogeneity and intrinsic variability of the coronal structure would lead to temporal variability of the coronal rain as 55\,Cnc e encounters different magnetic and energetic environments along its orbit and over time. Interestingly, the 41 days separating Visit B$_\mathrm{fuv}$ and C$_\mathrm{fuv}$ are close to the stellar rotation period ($\sim$40\,days; \citealt{Fischer2008}, \citealt{lopez2014}, \citealt{Demory2016a}). The strong similarities of the variations observed in these two visits (Sect.~\ref{sec:summary}) suggest that the stellar corona kept a similar structure over one stellar rotation and that we observed about the same stellar phase. As a result, the planet would have moved through similar regions of the corona in Visits B$_\mathrm{fuv}$ and C$_\mathrm{fuv}$. Because Visit A$_\mathrm{fuv}$ occurred nine months earlier, its different behavior can be interpreted as changes in the coronal structure and observed stellar phase.\\

The coronal rain scenario might also tentatively explain the successive absorption signatures seen in the \ion{C}{ii} doublet in Visit B$_\mathrm{fuv}$ (Sect.~\ref{sec:summary}). Absorption in the core and red wing of the \ion{C}{ii} ground-state line could trace ionized carbon atoms flowing down in the coronal rain along with N$^\mathrm{4+}$ and Si$^\mathrm{3+}$ ions, since the corresponding stellar lines show absorption at similar orbital phases. The absorption signature in the excited \ion{C}{ii} line, which has about the same shape as in the ground-state line but occurs during the following HST exposure and at negative velocities, might indicate that the ionized carbon atoms in the downflow were then excited and expelled back from the star. \\


\subsubsection{Similarities with the hot-Jupiter HD\,189733 b}

It is interesting to compare 55\,Cnc with the evaporating hot-Jupiter HD\,189733 b. Using the same wavelength range as our observations, \citet{Pillitteri2015} detected enhanced emission in multiple stellar chromospheric lines after the planet eclipse. The \ion{Si}{iv}, \ion{C}{iii,} and \ion{Si}{iii} lines of HD\,189733 showed the strongest enhancement, more than twice as strong as the \ion{N}{v} and \ion{C}{ii} lines, and even stronger than the \ion{Si}{ii} line. The three most enhanced lines are also those for which we detected a brightened emission in 55\,Cnc (Sect.~\ref{sec:summary} ), albeit with a much lower amplification factor (about 6\%, compared to $\sim$100-600\% for HD\,189733). The enhancement might thus have arisen from the same process in both systems, but was too faint in the case of 55\,Cnc to be detected in other chromospheric lines. \citet{Pillitteri2015} interpreted their observations as gas escaping from the hot Jupiter and funneled by the magnetized stellar wind onto the stellar surface at 70-90$^{\circ}$ ahead of the subplanetary point. The authors proposed that the resulting hot spot and the hot, dense plasma at the base of the stream cause the increased emission when they emerge at the stellar limb. In the simulations from \citet{Pillitteri2015} (see also \citealt{Matsakos2015}) the planetary outflow is the main contributor to the material in such an accretion stream. Yet it is not clear whether a hot Jupiter could lose silicon, carbon, and nitrogen gas in such amounts as required to explain the enhanced emission. Furthermore, one would expect this dense planetary outflow to yield significant absorption in the lines of ionized species before the planetary transit. A pre-transit absorption was indeed observed in the \ion{C}{ii} doublet (\citealt{BJ_ballester2013}), but deemed unlikely to originate from a planetary outflow or shocked stellar wind (\citealt{BenJaffel2014}). A circumstellar torus sustained by an Io-like moon (\citealt{BenJaffel2014}) or outgassing from Trojan satellites (\citealt{Kislyakova2016}) have instead been proposed as possible sources, yet cannot provide the amount of gas required to explain the absorption of HD\,189733 \ion{C}{ii}  doublet (\citealt{Kislyakova2016}). We thus wonder whether the enhanced emission and \ion{C}{ii} absorption in the chromospheric lines of HD\,189733 could be induced by the hot Jupiter into its star without the need for accreting planetary material, as proposed for the 55\,Cnc system. \citet{Pillitteri2015} mentioned similarities between their observations and the FUV spectrum of solar eruptions, in which filaments eject dense fragments that eventually fall down on the solar chromosphere. HD\,189733 b could destabilize the stellar corona through magnetic connections, leading to the observed flares and subsequent downfall of cooled coronal plasma onto the chromosphere. \\


\section{Conclusions}
\label{sec:conclu}

Our results join a growing list of observational hints of SPIs (see \citealt{Miller2015} for a review). \citet{France2016} interpreted enhanced L(\ion{N}{v})/L$_\mathrm{Bol}$ ratio as tentative observational evidence of planets with large planetary mass-to-orbital distance ratios interacting with the transition regions of their host stars. Despite its low mass, 55\,Cnc orbits close enough to its star that it yields a relatively large M$_\mathrm{p}$/a$_\mathrm{p}\sim$524\,M$_{\oplus}/$au. However, we measure L(\ion{N}{v})/L$_\mathrm{Bol}$ = 3.6$\times$10$^{-8}$, which is much lower than the enhanced value expected from SPI interactions (Fig.~16 in \citet{France2016}). This tracer thus provides no clear evidence for SPI in the 55\,Cnc system, and we do not expect that the influence of 55\,Cnc e on its host star can be easily detected except through high-resolution spectroscopy over short timescales. \\

Our spectrally resolved observations of 55\,Cnc chromospheric emission lines over three epochs hint at a complex case of SPIs. Visits B$_\mathrm{fuv}$ and C$_\mathrm{fuv}$ revealed repeatable variations with a pre-transit enhancement in the \ion{C}{iii}, \ion{Si}{iii,} and \ion{Si}{iv} lines followed by flux decreases in the \ion{Si}{iv} doublet. In Visit B$_\mathrm{fuv}$ this occurred simultaneously with decreases in the \ion{Si}{ii} and \ion{C}{ii} doublets. All visits further showed flux decreases in the \ion{N}{v} doublet, albeit at different orbital phases. On the one hand, the temporal and spectral properties of these signatures make them unlikely to arise from intrinsic stellar variability (Sect.~\ref{sec:unl_storig}). On the other hand, the flux decreases are explained well by absorption from optically thin gas clouds, but the irradiation and composition of 55\,Cnc e do not support a planetary origin for this gas (Sect.~\ref{sec:unlik_plorig}). We thus propose that the above variations are induced within the stellar atmosphere by 55\,Cnc e and that the orbital motion of the planet so close to the star triggers a cool coronal rain (Sect.~\ref{sec:cool_rain}). Plasma downflows would be first visible in emission and then in absorption as they would cool down from coronal to sub-chromospheric temperatures. In this scenario, the SPI and its observable signatures are strongly linked to the structure of the corona. The short time span between visits B$_\mathrm{fuv}$ and C$_\mathrm{fuv}$, which incidentally corresponds to about one stellar revolution, would explain the similarities between the two visits and their difference with Visit A$_\mathrm{fuv}$ obtained nine months earlier. We note that the coronal rain scenario fails to explain the change in the shape of the \ion{C}{ii} doublet over Visit A$_\mathrm{fuv}$, and the apparently random absorptions and distortions in the low-excitation \ion{O}{i} lines. These variations might thus have a different origin, which could be intrinsic stellar variability. Assessing the validity of our scenario and the origin of 55\,Cnc chromospheric variations will require full coverage of the planetary orbit at FUV wavelengths, and 3D magneto-hydrodynamical simulations of the stellar chromosphere/corona coupled with the planet. \\

The present study is part of a larger effort to investigate the 55\,Cnc system at UV wavelengths, but complementary observations in all wavelengths domains will be required to get the full picture of 55\,Cnc e environment, including analysis of the Rossiter-McLaughlin effect to derive the trajectory of the planet through the stellar magnetosphere (possibly misaligned; see \citealt{bourrier2014b}, \citealt{lopez2014}) and spectropolarimetric observations to map its 3D structure.  \\


\begin{acknowledgements}
We thank our referee for an encouraging review of this complex analysis and for suggestions. We deeply thank B.O. Demory for revising the ephemeris of 55\,Cnc e. We kindly thank P.A. Wilson for useful discussions about the airglow and for allowing us to use measurements from \citealt{Wilson2017}. We thank the STScI for their help with COS spectral calibration and LSFs. This work has been carried out in the frame of the National Centre for Competence in Research ``PlanetS'' supported by the Swiss National Science Foundation (SNSF). V.B. acknowledges the financial support of the SNSF. This project has received funding from the European Research Council (ERC) under the European Union's Horizon 2020 research and innovation program (project Four Aces grant agreement No 724427). A.L.E acknowledges support from the Centre National des Etudes Spatiales (CNES). CHIANTI is a collaborative project involving George Mason University, the University of Michigan (USA), and the University of Cambridge (UK).
\end{acknowledgements}

\bibliographystyle{aa} 
\bibliography{biblio} 

\begin{appendix}

\section{Far-ultraviolet  lines of 55\,Cnc}
\label{apn:full_FUV_lines}

\begin{figure*}
\centering
\begin{minipage}[b]{\textwidth}
\includegraphics[trim=0cm 0cm 0cm 0cm,clip=true,width=\columnwidth]{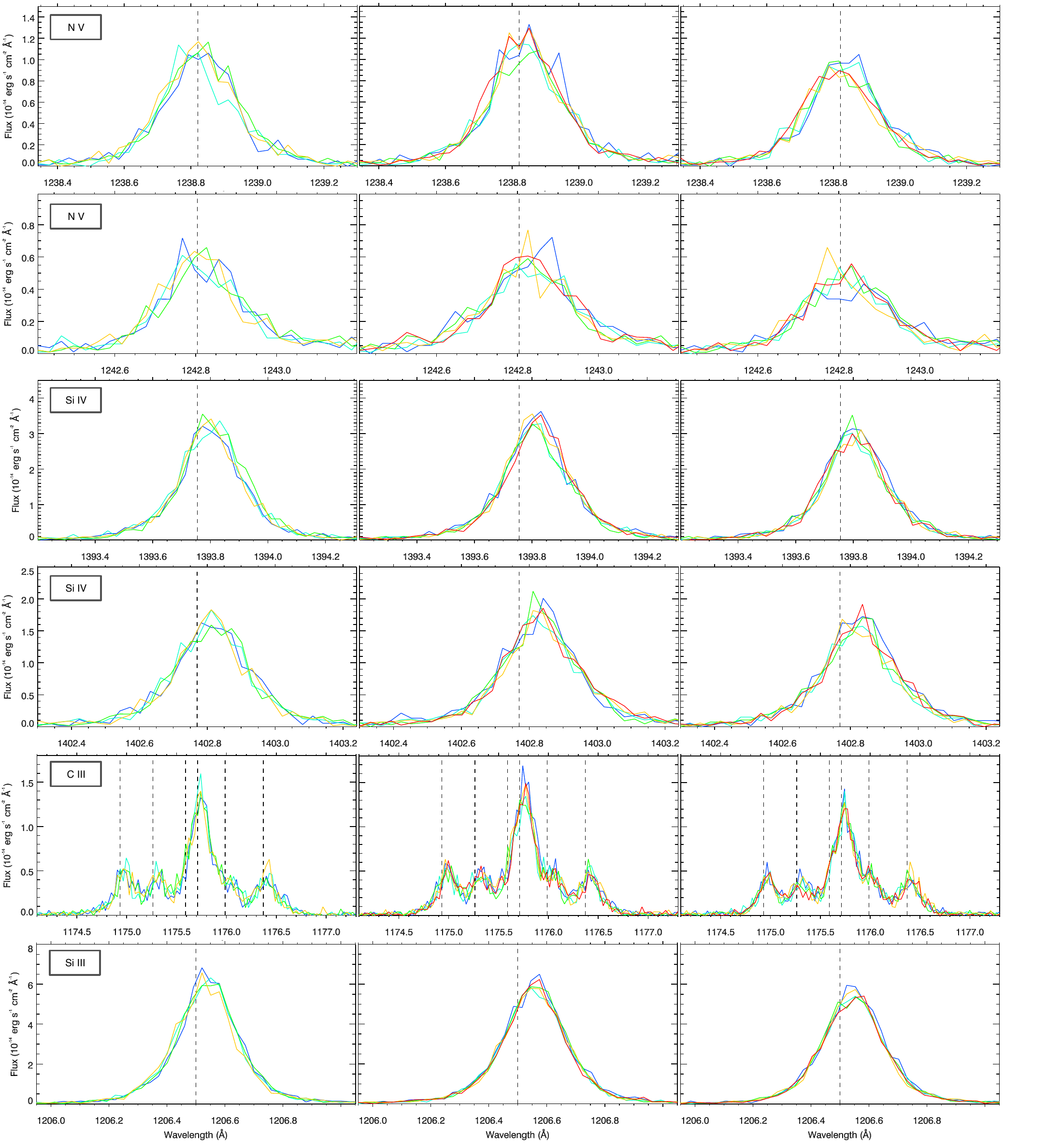}
\end{minipage}
\caption[]{Spectral profiles of 55\,Cnc FUV lines, uncorrected for their spectral shifts. Vertical dashed lines correspond to the rest wavelength of the stellar lines in the expected star rest frame. Each row corresponds to a stellar line. First to third columns correspond to visits A$_\mathrm{fuv}$, B$_\mathrm{fuv}$, and C$_\mathrm{fuv}$, respectively. Spectra in exposures 0 to 4 are indicated in blue, cyan, green, orange, and red, respectively (the first visit has no exposure 4).} 
\label{fig:Lines_spec_grid_raw}
\end{figure*}

\begin{figure*}\ContinuedFloat
\centering
\begin{minipage}[b]{\textwidth}
\includegraphics[trim=0cm 0cm 0cm 0cm,clip=true,width=\columnwidth]{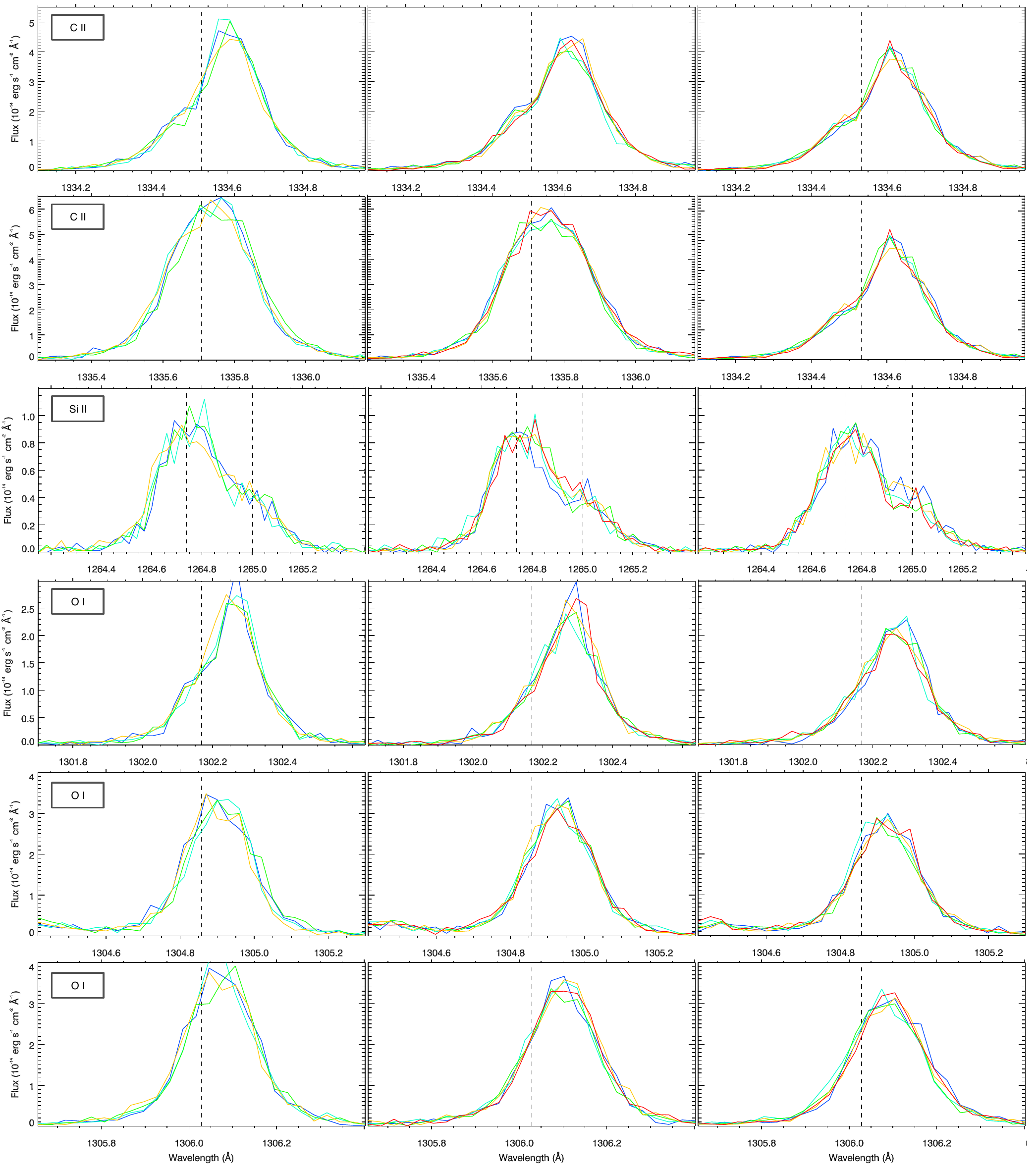}
\end{minipage}
\caption[]{Continued}
\label{fig:Lines_spec_grid_raw}
\end{figure*}

\end{appendix}

\end{document}